\documentclass[traditabstract]{aa}  
\usepackage{graphicx}

\usepackage{txfonts}
\newcommand{\s}{{\it Spitzer}}
\newcommand{\h}{{\it Herschel}}
\newcommand{\msol}{M$_{\odot}$}

\newcommand{\lsol}{L$_{\odot}$}
\newcommand{\lir}{$L_{\rm IR}$}
\newcommand{\pacs}{$S_{\rm 100}$}
\newcommand{\mips}{$S_{\rm 24}$}

\newcommand{\irs}{$S_{\rm 16}$}
\newcommand{\pam}{$S_{\rm 100}$/$S_{\rm 24}$}
\newcommand{\irm}{$S_{\rm 16}$/$S_{\rm 24}$}
\newcommand{\iri}{$S_{\rm 16}$/$S_{\rm 8}$}
\newcommand{\temp}{$S_{\rm 160}$/$S_{\rm 100}$}
\begin{document}

\title{GOODS-{\it Herschel}: A population of $24\,\mu$m dropout sources at $z$ $<$ 2
\thanks{Herschel is an ESA space observatory with science instruments provided
  by European-led Principal Investigator consortia and with important participation from NASA.}}
\author{Georgios E. Magdis\inst{1,2}
\and D.~Elbaz \inst{1}
\and M.~Dickinson\inst{4}
\and H.S.~Hwang\inst{1}
\and V.~Charmandaris\inst{5,6,7}
\and L.~Armus\inst{8}
\and E.~Daddi\inst{1}
\and E.~Le Floc'h\inst{1}
\and H.~Aussel\inst{1}
\and H. Dannerbauer\inst{1}
\and D. Rigopoulou \inst{2,3}
\and V.~Buat\inst{9}
\and G. Morrison\inst{10}
\and J.~Mullaney\inst{1}
\and D.~Lutz\inst{11}
\and D.~Scott\inst{12}
\and D.~Coia\inst{13}
\and A.~Pope\inst{14}
\and M.~Pannella\inst{1}
\and B.~Altieri\inst{13}
\and D.~Burgarella\inst{9}
\and M.~Bethermin\inst{15}
\and K.~Dasyra\inst{1}
\and J.~Kartaltepe\inst{4}
\and R.~Leiton\inst{1}
\and B.~Magnelli\inst{11}
\and P.~Popesso\inst{11}
\and I.~Valtchanov\inst{13}
}

\institute{ Laboratoire AIM-Paris-Saclay, CEA/DSM/Irfu - CNRS - Universit\'e Paris Diderot, CE-Saclay, F-91191 Gif-sur-Yvette, France \\
\email{georgios.magdis@astro.ox.ac.uk}
\and Department of Physics, University of Oxford, Keble Road, Oxford OX1 3RH
\and Space Science \& Technology Department, Rutherford Appleton Laboratory, Chilton, Didcot, Oxfordshire OX11 0QX
\and National Optical Astronomy Observatory, 950 North Cherry Avenue, Tucson, AZ 85719, USA
\and Department of Physics \& Institute of Theoretical and Computation Physics, University of Crete, GR-71003, Heraklion, Greece
\and IESL/Foundation for Research \& Technology-Hellas, GR-71110, Heraklion, Greece
\and Chercheur Associ\'e, Observatoire de Paris, F-75014,  Paris, France
\and Spitzer Science Center, California Institute of Technology, Pasadena, CA.
\and Laboratoire d'Astrophysique de Marseille, OAMP, Université Aix-Marseille, CNRS, 38 rue Frédéric Joliot-Curie, 13388 Marseille Cedex 13
\and Institute for Astronomy, University of Hawaii, Manoa, HI 96822, USA; Canada-France-Hawaii Telescope Corp., Kamuela, HI 96743, USA
\and Max-Planck-Institut f\"ur Extraterrestrische Physik (MPE), Postfach 1312, 85741, Garching, Germany
\and Department of Physics and Astronomy, University of British Columbia, Vancouver, BC V6T 1Z1, Canada
\and Herschel Science Centre, European Space Astronomy Centre, Villanueva de la Ca\~nada, 28691 Madrid, Spain
\and Department of Astronomy University of Massachusetts, LGRT-B 618, 710 North Pleasant Street, Amherst, MA 01003
\and Univ Paris-Sud, Laboratoire IAS, UMR8617, 91405, Orsay Cedex, France; CNRS, Orsay, 91405, France
}


   
\abstract{
Using extremely deep PACS 100- and $160\,\mu$m \h ~data from the GOODS-\h ~program, we identify 21 infrared bright galaxies previously missed in the deepest $24\,\mu$m surveys performed by \s /MIPS. These MIPS dropouts are predominantly found in two redshift bins, centred at $z$ $\sim$0.4 and $\sim$1.3. Their \pacs/\mips ~flux density ratios are similar to those of local (ultra-) luminous infrared galaxies (LIRGs and ULIRGs), whose silicate absorption features at $18\,\mu$m (at $z$ $\sim$ 0.4) and $9.7\,\mu$m (at $z$ $\sim$ 1.3)  are shifted into the $24\,\mu$m MIPS band at these redshifts. The high-$z$ sub-sample consists of  11 infrared luminous sources, accounting for $\sim$ 2\% of the whole GOODS-\h ~sample and putting strong upper limits on the fraction of LIRGs/ULIRGs at 1.0 $<$ $z$ $<$ 1.7 that are missed by the $24\,\mu$m surveys. We find that a \pacs/\mips ~$>$ 43 colour cut selects galaxies with a redshift distribution similar to that of the MIPS dropouts and when combined with a second colour cut, \iri ~$>$ 4, isolates sources at 1.0 $<$ $z$ $<$ 1.7. We show that these sources have elevated specific star formation rates (sSFR) compared to main sequence galaxies at these redshifts and are likely to be compact starbursts with moderate/strong $9.7\,\mu$m silicate absorption features in their mid-IR spectra. \h ~data reveal that their  infrared luminosities  extrapolated from the $24\,\mu$m flux density are underestimated, on average, by a factor of $\sim$ 3. These {\it silicate break} galaxies account for 16\% (8\%) of the ULIRG (LIRG) population in the GOODS fields, indicating a lower limit in their space density of 2.0 $\times$ 10$^{-5}$ Mpc$^{-3}$. Finally, we provide estimates of the fraction of $z<2$ MIPS dropout sources as a function of the 24-, 100-, 160-, 250- and $350\,\mu$m sensitivity limits, and conclude that previous predictions of a population of silicate break galaxies missed by the major $24\,\mu$m extragalactic surveys have been overestimated.}


\titlerunning{GOODS-H; $24\,\mu$m drop-out galaxies}
\authorrunning{G. E. Magdis et al.}
\maketitle

\section{Introduction}
Accurate measurement of star formation rates (SFR) is a key ingredient for studying galaxy evolution and deriving the census
of the star formation activity, both in the distant and in the local universe.
To this end, it has been shown that the contribution of  luminous infrared 
galaxies (with infrared luminosities \lir ~$>$ 10$^{11}$ \lsol) to the star formation density 
is progressively rising as we look back in cosmic time, at least up to $z$ $\sim$ 2. Indeed, 
although they were found to be rare in the local Universe and to account for only $\sim$ 5$\%$ of 
the total infrared energy emitted by galaxies at $z$ $\sim$ 0 (Soifer et al. 1991, Kim \& Sanders 1998), LIRGs and ULIRGs (\lir ~$>$ 10$^{12}$ \lsol), dominate the SFR density at $z$ $\sim$ 1$-$2, accounting for 70$\%$ 
of the star formation activity at these epochs (Papovich et al. 2004, Le Floc'h et al. 2005, Caputi et al. 2007).

The study of infrared sources was greatly facilitated by the advent of the Spitzer Space Telescope (\s ~Werner et al. 2004). Extragalactic surveys carried out using the MIPS $24\,\mu$m band on-board \s ~confirmed the strong evolution of 
these sources first indicated by infrared and submillimeter observations using {\it ISO} and SCUBA, 
respectively (Blain et al. 1999a, Elbaz et al. 1999, Serjeant et al. 2001, Dole et al. 2001). Such surveys are believed to detect the bulk of the dusty star forming galaxies up to $z$ $\sim$2. However there are two important caveats. The first is that the conversion of $24\,\mu$m flux densities to total \lir ~(and therefore SFR) is subject to large uncertainties, as it relies on extrapolations that strongly depend on the assumed spectral energy distribution (SED) libraries (Chary \& Elbaz 2001; Lagache et al. 2003; Dale \&  Helou 2002). The second comes from the prominent emission and absorption features between 3- and $19\,\mu$m in the spectra of star forming galaxies 
that parade through the $24\,\mu$m band at various redshifts.

A large number of studies using the Infrared Spectrograph (IRS, Houck et al. 2004) have revealed that the vast majority of local LIRGs and ULIRGs exhibit a broad silicate absorption feature centred at $9.7\,\mu$m with silicate optical depths ranging from $\tau_{9.7}$ $\sim$ 0.4 to $\tau_{9.7}$ $\geq$ 4.2 (Brandl et al. 2006; Armus et al. 2007; Pereira-Santaella et al. 2010). Furthermore, using a sample of local ULIRGs, Desai et al. (2007) found strong PAHs and prominent silicate absorption in the \ion{H}{II} and LINER sources and weak PAHs and silicate absorption in Seyferts, suggesting that ULIRGs with strong PAHs but weak silicate absorption are rare. Similar features have been observed in the mid-IR spectra of high-$z$ galaxies (Higdon et al. 2004, Houck et al. 2005). For example, Men{\'e}ndez-Delmestre et al. (2009) and Farrah et al. (2008) report a median $\tau_{9.7}$ $\sim$ 0.31 for a sample of submillimetre galaxies (SMGs) and IRAC selected ULIRGs respectively, while Sajina et al. (2007) found deeper silicate absorption features ($\tau_{9.7}$ $>$ 1.1) in a sample of $z$ $\sim$ 2 radio-loud galaxies. Although for LIRGs and ULIRGs the identification and measurement of the silicate optical depths is straightforward, this is not the case for normal galaxies (\lir ~$<$ 10$^{10}$ \lsol), as for the latter, it is difficult to discriminate between moderate PAH emission superimposed on a silicate-absorbed continuum and strong PAH features with a relatively weak underlying continuum (Smith et al. 2007). 

Whatever its origin, the existence of this broad  dip in the mid-IR spectra of star forming galaxies $10\,\mu$m could be of particular importance for galaxies in the redshift range of 1 $<$ $z$ $<$ 1.8. At these redshifts the $24\,\mu$m filter samples this part of the spectrum and sources with such features 
would appear faint at $24\,\mu$m or even be undetected in this band (depending on the depth of the $24\,\mu$m data). A second  broad dip that is common in the spectra of star forming galaxies is caused by another silicate absorption feature at $18\,\mu$m. This feature would have a similar effect for sources at 0.2 $<$ $z$ $<$ 0.6.

The impact of these features on the mid-IR colours as a function of redshift 
were presented in detail by Takagi \& Pearson (2005), who predicted a population of infrared 
luminous galaxies at $z$ $\sim$ 1.5 which, due to strong absorption at $9.7\,\mu$m, are not detected in the $24\,\mu$m band. 
Subsequent studies that focussed on the search for such silicate absorbed systems employed the $16\,\mu$m IRS peak-up image. 
Kasliwal et al. (2005) suggested that such objects account for more than half of all the 
sources at $z$ $\sim$ 1$-$2 predicted by various models. It has also been proposed that 
the mid-IR colour anomalies introduced by the silicate absorption feature can serve as a 
redshift indicator for dusty infrared luminous galaxies at $z$ $\sim$1.5 
(Charmandaris et al. 2004, Teplitz et al. 2011, Armus et al. 2007). Similar claims have also been presented by Pearson et al. (2010), using the AKARI IRC L18W to MIPS24 band colour. 
These studies raised concerns about a possible bias introduced by the $24\,\mu$m selection, 
in the sense that a significant fraction of $z \leq 2$ LIRGs and ULIRGs could remain undetected in $24\,\mu$m surveys. However, with little or no information about the far-IR part of the spectrum, 
these studies were subject to large extrapolation and hence suffered from large uncertainties. 

With the successful launch of the Herschel Space Observatory 
({\it Herschel}, Pilbratt et al. 2010), we now have access to wavelengths that 
directly probe the peak of the far-IR emission of high-$z$ galaxies and are 
in a position to measure with
 unprecedented accuracy their bolometric output. Deep \h ~extragalactic surveys can be used to determine the accuracy of our extrapolations of the far-IR properties of high-$z$ galaxies as well as test previous claims that $24\,\mu$m surveys miss a population 
 of $z < 2$ LIRGs and ULIRGs ($24\,\mu$m dropouts). In this paper, we use the deepest \h ~
 observations to date, as part of the GOODS-\h ~(GOODS-H) program (PI D. Elbaz), 
 covering both the north and the south part of the GOODS fields (GOODS-N and GOODS-S 
 respectively), (Dickinson et al. 2003, Giavalisco et al. 2004), to search for such sources. In Section 2 we present the Herschel data, introduce the GOODS-H sample of galaxies, and identify $24\,\mu$m dropout sources, i.e., sources detected in the PACS bands but not at $24\,\mu$m. In Section 3 we investigate the properties of this population, while in Section 4 we extend our study to the whole GOODS-H sample. Finally in Section 5 we provide estimates of the fraction of $z$ $<$2 MIPS dropout sources as a function of the $24\,\mu$m, $100\,\mu$m and $160\,\mu$m sensitivity limits and summarize our results.
 
\section{Herschel data and sample selection}

\h ~observations were obtained as part of the open
time key program GOODS-H (PI D.Elbaz). The full $10^{\prime}$ $\times$ $16^{\prime}$ GOODS-N field was imaged 
with the PACS (Poglitsch et al. 2010) and SPIRE (Griffin et al. 2010) instruments
at 100, $160\,\mu$m (PACS) and 250, 350, $500\,\mu$m (SPIRE). 
The total observing time was 124.6 hours ($\sim$ 2.5h / sky position) and 31.1 hours  for PACS and SPIRE respectively. 
Similarly a $7^{\prime}$ $\times$ $7^{\prime}$ part of the GOODS-S field was observed by PACS over a total of 264
  hours ($\sim$ 15h / sky position). Observations of both fields were carried out by adopting the intermediate speed ($20^{\prime\prime} s^{-1}$) 
  scan-map mode. Both PACS and SPIRE data were processed through the standard \h ~reduction pipeline, 
  version 6.0.3, within the HCSS environment. Additionally, we employed 
  custom procedures aimed at removing of interference patterns, tracking anomalies, re-centering positional offsets, and mapping. A full description of the data reduction procedures will be given in
 a companion paper (Leiton et al. 2011 in prep).

\subsection{Prior based Source extraction; The GOODS-H sample}
Given the large beam size of the \h ~bands, (FWHM $\sim$ 6.7$^{\prime\prime}$, 11.2$^{\prime\prime}$,18.0$^{\prime\prime}$, 25.0$^{\prime\prime}$, 36.0$^{\prime\prime}$ for
 PACS 100- and $160\,\mu$m and SPIRE 250, 350 and $500\,\mu$m), a common approach to performing source extraction has been 
 a guided extraction using priors. Here we will give a brief summary of the procedure as an extensive description of the method is given in Elbaz et al. (2011). Source extraction and photometry were obtained from point source fitting at prior positions defined by $24\,\mu$m sources with fluxes brighter than \mips ~$\sim$ 20 $\mu$Jy for the $100\,\mu$m maps and down to \mips ~$\sim$ 30 $\mu$Jy for the $160\,\mu$m and $250\,\mu$m maps. For the other two SPIRE bands, a secondary criterion was needed, as the $24\,\mu$m sources were far too numerous and would lead to an over-deblending of the actual sources. Hence, only sources with S/N $>$ 2 at $250\,\mu$m were considered as priors for the longer wavelength  SPIRE bands. This choice was optimized by Monte Carlo (MC) simulations to avoid artificial over-deblending of a source, but also to give clean residual maps.
\begin{figure*}
\centering
\includegraphics[scale=0.7]{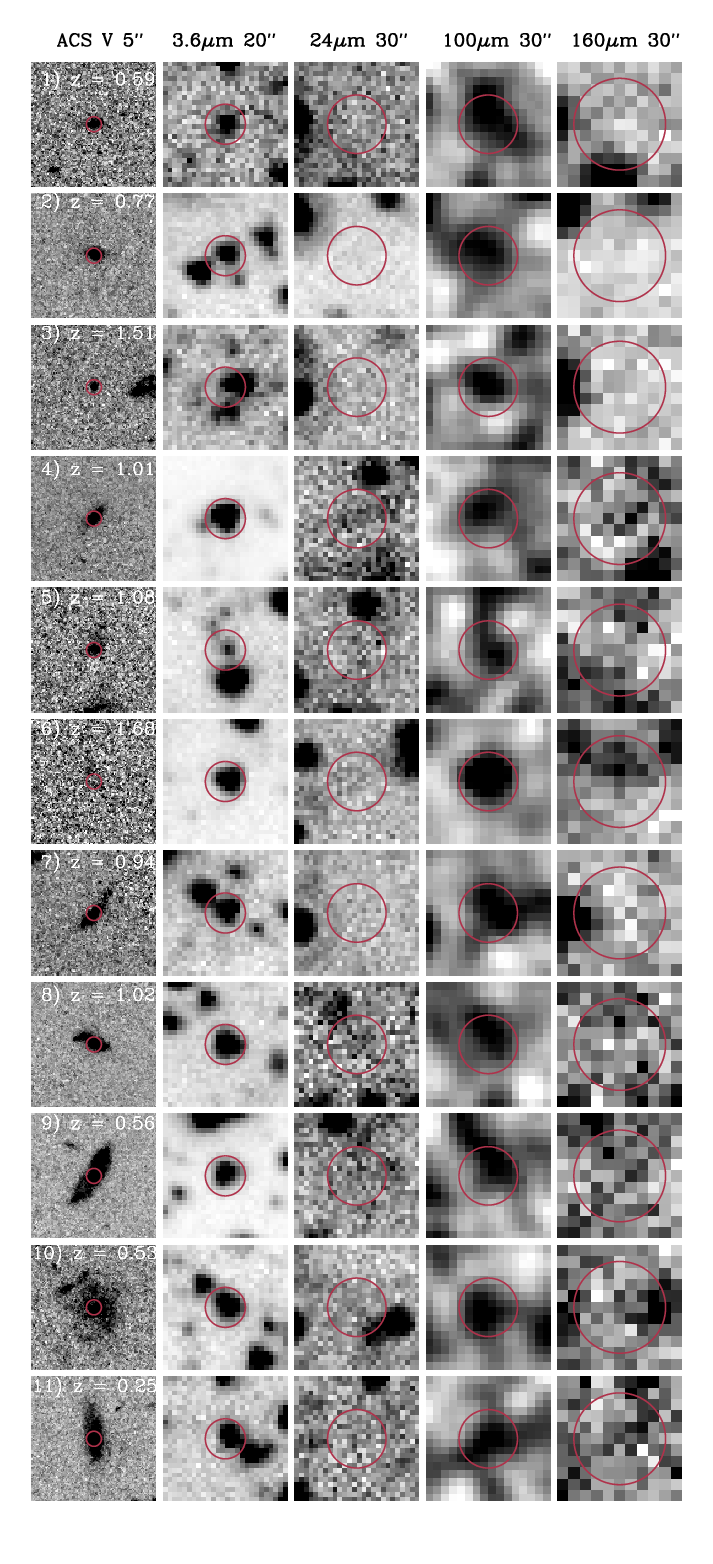} 
\includegraphics[scale=0.7]{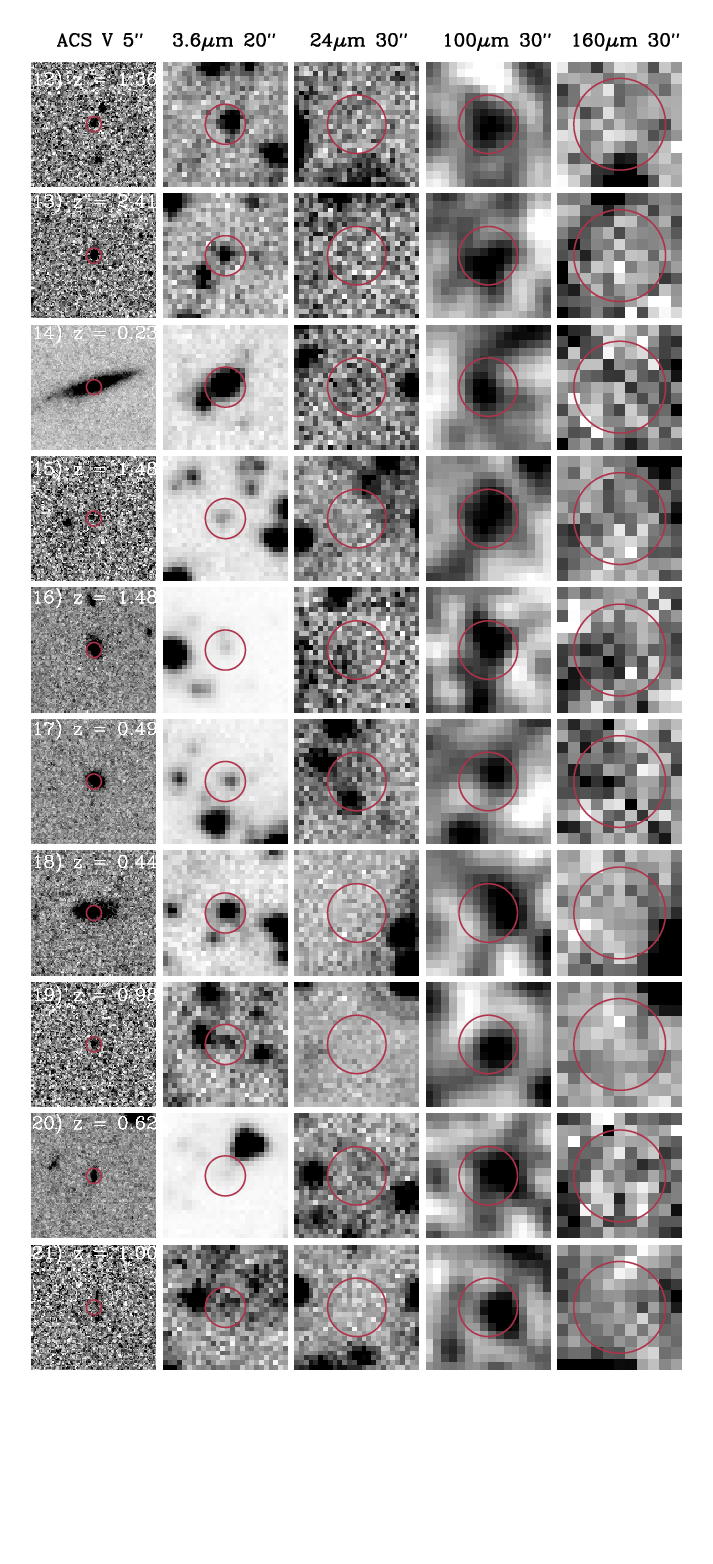}\\
\caption{\small{ACS $V-$band (5'' $\times$ 5''), IRAC $3.6\,\mu$m (20'' $\times$ 20''), $MIPS 24\,\mu$m (30'' $\times$ 30''), and PACS 100-, 160$\,\mu$m (30'' $\times$ 30'') cut-out images of the 24$\,\mu$m dropout sources from our sample.
The red circles are centred at the IRAC $3.6\,\mu$m positions of the sources and their diameter corresponds to the FWHM at each band. The size of each image is denoted on the top of each column.}}
\label{fig:sfsed}
\end{figure*}
Flux uncertainties were based on local estimates of the background noise at the position of the sources from residual images produced after subtracting detected sources, while global noise estimates for the maps were derived from Monte Carlo simulations. 
This used artificial sources injected into the \h ~maps 
and source extraction performed in the same manner as for the real sources. The dispersion
between input and recovered fluxes provides a secure estimate of the completeness and the noise properties of the map. The two noise estimates were found to be in good agreement.

To construct the GOODS-H sample, we considered sources with flux densities down to 3$\sigma$ in the PACS bands, i.e. 1.0 and 2.6 mJy (0.7 and 2.6 mJy) at 100 and $160\,\mu$m in GOODS-N (GOODS-S). For the GOODS-N sample, where SPIRE data are available we also considered sources down to the 5$\sigma$ detection limit, i.e. 6.3, 7.1, and 15.0 mJy at 250, 350 and $500\,\mu$m, respectively. The choice of a higher S/N cut for the SPIRE catalogues was dictated by the larger beam size and the confusion noise that significantly affects the SPIRE observations (for more details see Elbaz et al. 2011). 
Herschel catalogs were then matched with the existing multi-wavelength data of the GOODS team to create a multi-band merged catalogue of GOODS galaxies including {\it HST ACS BViz, J,K, IRAC, Spitzer MIPS} 24 and $70\,\mu$m, Herschel PACS 100 and $160\,\mu$m, and \h ~SPIRE 250, 350 and $500\,\mu$m. Among our sources, 65$\%$ have secure spectroscopic redshifts, while for the rest, we use the reliable compilation of 
photometric redshifts by Le Borgne et al (2009). Hereafter we will refer to this sample as the GOODS-H sample.

\subsection{Blind source extraction; The $24\,\mu$m dropout sample}
Since the main aim of this work is to investigate whether \h ~observations reveal a population of
 galaxies that were previously missed by $24\,\mu$m surveys, we also performed blind source extraction in the two PACS bands using {\it Starfinder}, a point spread function (PSF) fitting code (Diolaiti et al. 2000). We first extracted PSF profiles from the final science maps that were used to perform source extraction. Aperture corrections were derived based on calibration observation of the asteroid Vesta, while the flux uncertainties were derived based both on the error maps and Monte Carlo simulations, as described above. Monte Carlo simulations were also employed to obtain the level of completeness and the fraction of spurious sources. Both the derived fluxes and the noise properties of the maps are in good agreement with those obtained by the prior based source extraction. Finally, a critical parameter in  {\it Starfinder} is the {\it correlation threshold} (ct), a measure of the similarity between the PSF used for source extraction and the profile of the extracted source, with ct=1 corresponding to identical profiles. 
 MC simulations indicate that high ct values result in catalogs immune to spurious detections but with lower completeness levels. Similarly lower ct values correspond to higher completeness 
 but also to higher fractions of spurious sources. For our blind catalogs we  consider sources with flux densities above the 4$\sigma$ detection limit at each band and ct $>$ 0.67, for which the fraction of spurious sources is $<$ 3\%.

In principle, our aim was to find sources detected in either of the PACS bands but undetected at $24\,\mu$m. Therefore, we first matched the PACS 100- and $160\,\mu$m blind catalogs with the MIPS$24\,\mu$m sample down to \mips ~$\sim$20 $\mu$Jy (3$\sigma$), i.e. the one that served as a pool for the prior based source extraction, starting from the longest wavelength available and using search radii of $7^{\prime\prime}$ and $11^{\prime\prime}$ respectively. Sources with $24\,\mu$m counterparts were omitted while the rest were matched to the IRAC 3.6 $\mu$m catalogue and subsequently to the master GOODS multi-bands catalogue described in the previous section. We also performed photometry in the $24\,\mu$m maps at the position of the PACS sources to ensure that there is no $24\,\mu$m source at this position, possibly  missing from the $24\,\mu$m catalogue. We calculated the corrected Poissonian probability that an association of a PACS source within the search radius is a chance coincidence (see Downes et al. 1986) and  all sources were found to have a robust ($p$ $<$ 0.05) 3.6 $\mu$m counterpart. We also note that flux boosting due to insufficient de-blending, which is the main caveat of blind source extraction, should not, by definition, be an issue for the $24\,\mu$m dropout sample. 
All sources were also inspected by eye and a quality flag was  attributed to them. In particular, sources with multiple IRAC counterparts within the PACS beam and sources close to bright objects in the PACS bands were flagged as low quality sources. The final sample consists of 21 MIPS dropout sources, all detected at $100\,\mu$m and two at $160\,\mu$m, accounting for $\sim$ 2\% of the total sources detected in the PACS bands. Hence, we find that even at the confusion limit of the 100- and $160\,\mu$m passbands  (0.7- and 2.6 mJy at a 3$\sigma$ level), 98 \% of the \h ~sources possess a robust $24\,\mu$m counterpart brighter than 20 $\mu$Jy.  The 21 MIPS dropout sources are shown in Fig. 1 where we present cut$-$out images at several bands.
\section{$24\,\mu$m dropout sources}
The small number of $24\,\mu$m dropouts indicates that the vast majority of PACS sources do have a 
$24\,\mu$m counterpart. 
In other words, the ``normal" SED behaviour of galaxies in the GOODS sample is the one 
where the relative sensitivity at $24\,\mu$m overpowers that of the PACS bands. In Fig. 2 (left) we plot the 
flux density at $24\,\mu$m over that at $100\,\mu$m for the whole GOODS-H sample, as well as for the dropout 
sources and see that the latter depart from the general trend and are relatively faint at 100$\,\mu$m. 
We also note that similarly to the MIPS dropouts, some sources with $24\,\mu$m detection tend to 
exhibit redder \pam ~colours than the bulk of the population while they span a wide range of \mips. 
Here we will study the origin of the departure of the MIPS dropouts 
from the bulk of the GOODS-H sample and our investigation will be driven by their property that intrigued our interest in the first place i.e. their unusual \pacs/\mips ~colour. 

\subsection{Far-IR properties}
The total infrared luminosity (\lir ~= $L_{\rm8-1000\mu m}$) of galaxies 
in the sample was determined from the $100\,\mu$m flux density using the templates of (Chary \& Elbaz 2001, CE01) and Dale \& Helou (2002) (DH02). 
Despite the lack of data points at longer wavelengths, we note that 
the monochromatic derivations of total IR luminosities 
from the $100\,\mu$m flux density tend to be robust up to $z$ $\sim$ 1.5 (Elbaz et al. 2010). 
For the two sources with $160\,\mu$m detection \lir ~was determined
from the best fit of the two PACS points, using the whole library of SED templates
from CE01 independently of their luminosity (i.e. allowing normalization of all 
SEDs to the observations) as well as the DH02 templates. The derived luminosities range 
from 8 $\times$ 10$^9$ to 2 $\times$ 10$^{12}$ \lsol , with 11 of the MIPS dropout sources having 
\lir ~$>$10$^{11}$ \lsol, belonging to the class of luminous infrared galaxies.
 \begin{figure*}[!t]
\centering
\includegraphics[scale=0.28]{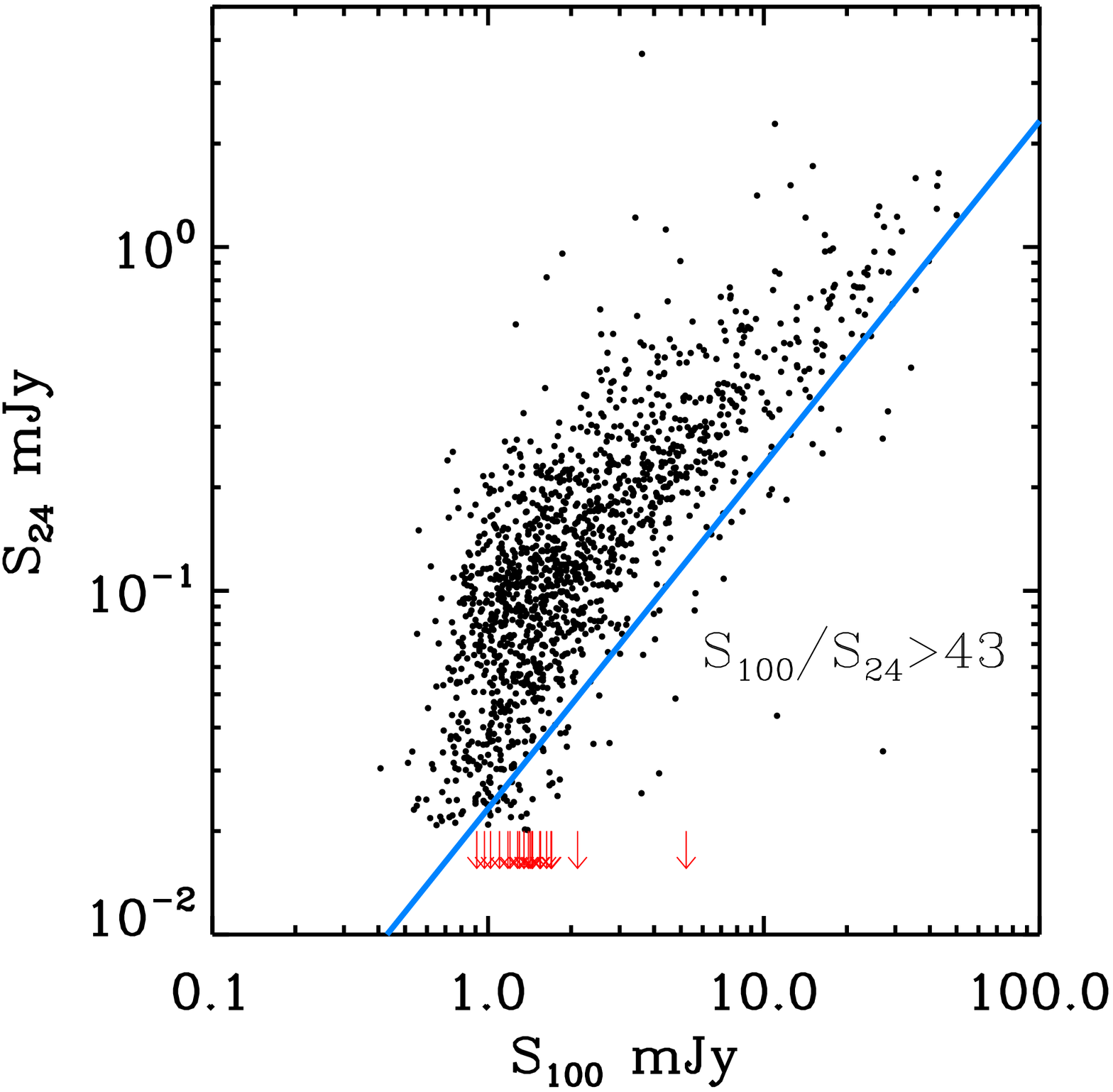}
\includegraphics[scale=0.28]{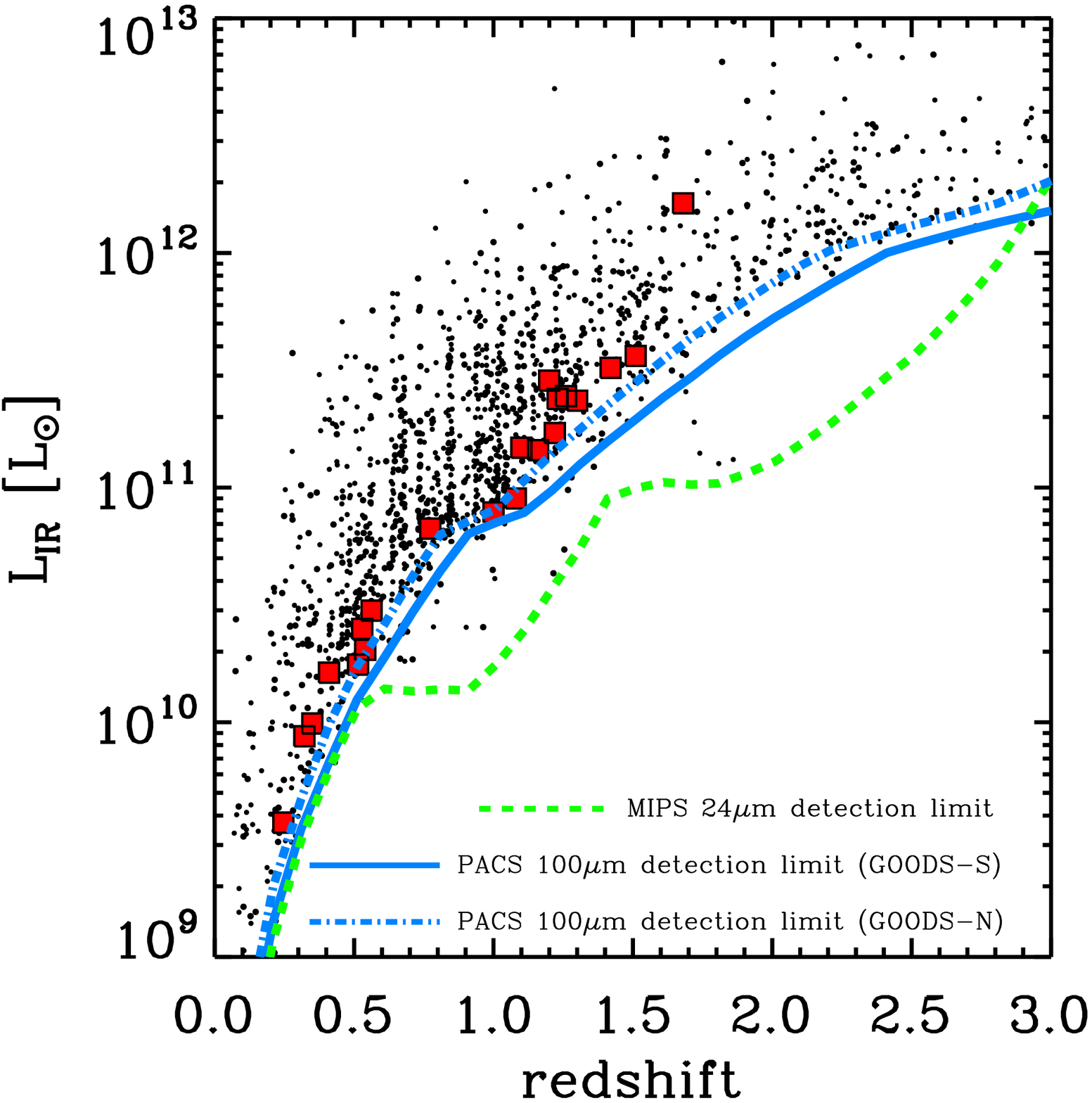}
\includegraphics[scale=0.28]{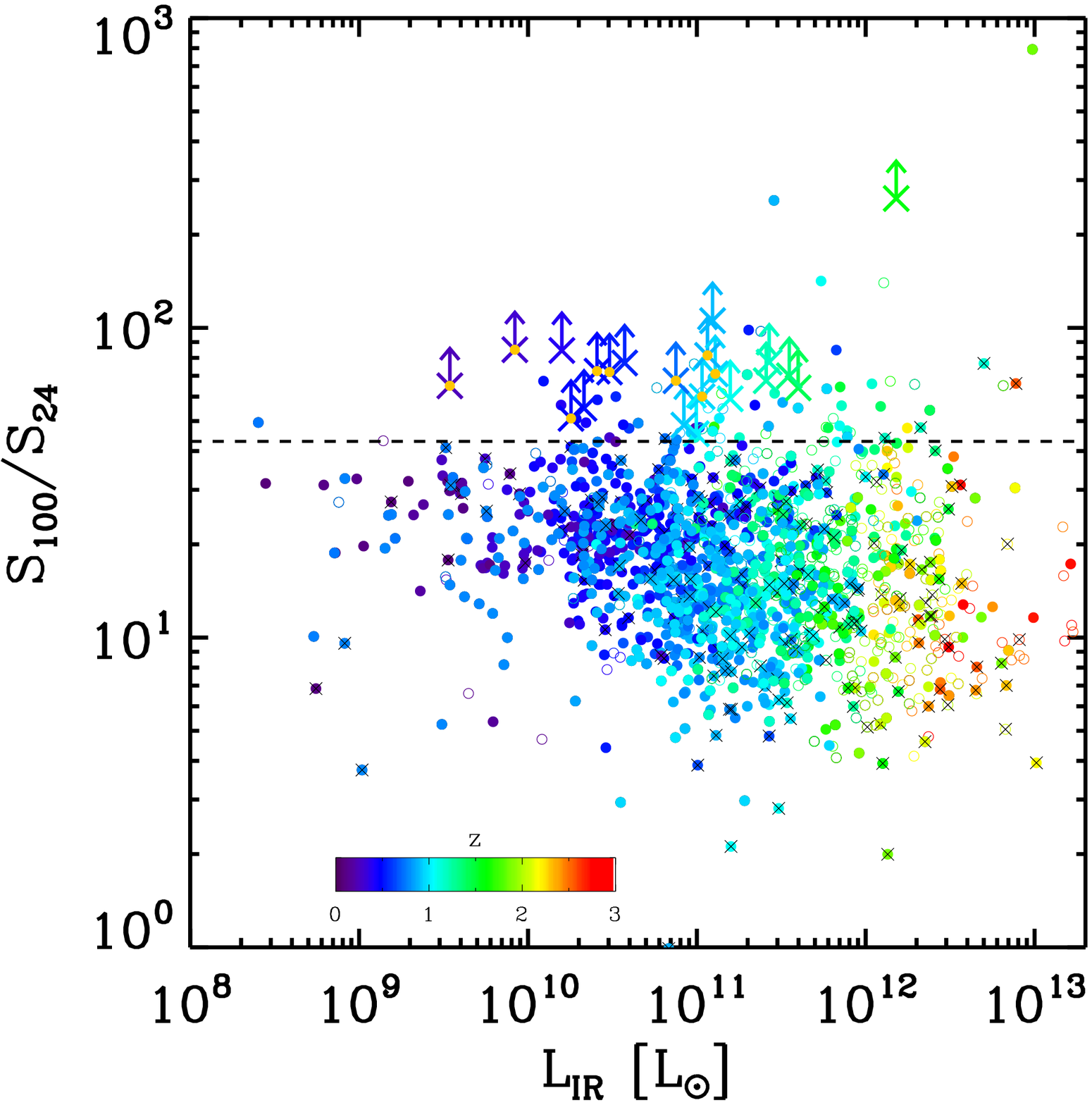}\\
\caption{Left: \mips ~vs \pacs~ flux densities for the whole GOODS-H sample (black circles) as well as for the $24\,\mu$m dropout sources (red arrows). For the dropouts we consider a 3 $\sigma$ upper limit for the \mips. Dropouts as well as some $24\,\mu$m detected sources tend to depart from the bulk of the GOODS-H population, exhibiting redder \pam ~ colours. The cyan line corresponds to \pam ~= 43.  As discussed latter in the paper, sources with  \pam ~$>$ 43 are classified as {\it silicate-break} galaxies. Middle: Detection limits  as a function
of redshift for the GOODS-N and GOODS-S PACS 100$\,\mu$m and  MIPS $24\,\mu$m observations. Red squares correspond to the drop-out sources. Right: \pam ~as a function of \lir ~as derived by \h ~for the whole GOODS-H sample (circles) and lower limits for the $24\,\mu$m dropouts (arrows). Both samples are colour coded based on their redshift. Sources with a black cross are  AGNs based on their X-ray emission. Filled symbols denote sources in the GOODS-H sample with spectroscopic redshifts while open symbols sources with photometric redshift. Similarly, yellow circles on top of the arrows indicate that a spectroscopic redshift is available for that $24\,\mu$m dropout source. The horizontal black dashed line corresponds to \pam ~= 43. }
\label{fig:sfsed}
\end{figure*}
 To illustrate the completeness of our sample in terms of \lir,  we plot the \lir ~ for the GOODS-H sample, as well as for the drop-out sources as a function of redshift, along with the corresponding detection limits at 100$\,\mu$m and 24$\,\mu$m (Fig.2 middle). Furthermore, In Fig. 2 (right), we also plot the \pacs/\mips ~colour as a function of \lir ~as derived from the \h ~data, both for the whole GOODS sample as well as for the MIPS dropouts (crosses with arrows). For the latter we compute lower limits to \pam ~assuming
 the 3$\sigma$ detection limit of the $24\,\mu$m maps.  The points are also colour$-$coded based on their redshift. We see that $24\,\mu$m dropouts have \lir ~values similar to that of the whole GOODS-H sample (for a given redshift), while they exhibit significantly higher \pam ~colours for the whole range of luminosities. As we discussed above, \lir ~scales with \pacs. Hence, given the richness of features in the rest frame MIR emission (i.e. Armus et al. 2007), it is more likely that the dropouts have a suppressed \mips ~emission rather than an excess at \pacs ~when compared to the rest of the GOODS-H sample. In what follows we investigate the origin of this \mips ~deficit. 
\subsection{\pam ~colour and redshift distribution}
We wish to examine whether the MIPS dropout sources tend to be found at specific redshifts. 
We note that nine sources have spectroscopic redshift, while for the rest we adopt the photometric redshifts derived by the GOODS team. It appears that $24\,\mu$m dropouts are distributed in two redshift bins, one centred at $z$ $\sim$ 0.4 and one at $z$ $\sim$ 1.3 (Fig. 3). This bimodality is verified by a KMM test (Ashman et al. 1994) at a 4.3$\sigma$ confidence level. Therefore, it seems that our sample is populated by low-$z$ (0.2 $<$ $z$ $<$ 0.6) and high-$z$ (0.9 $<$ $z$ $<$ 1.7) sources.
 \begin{figure}
\centering
\includegraphics[scale=0.4]{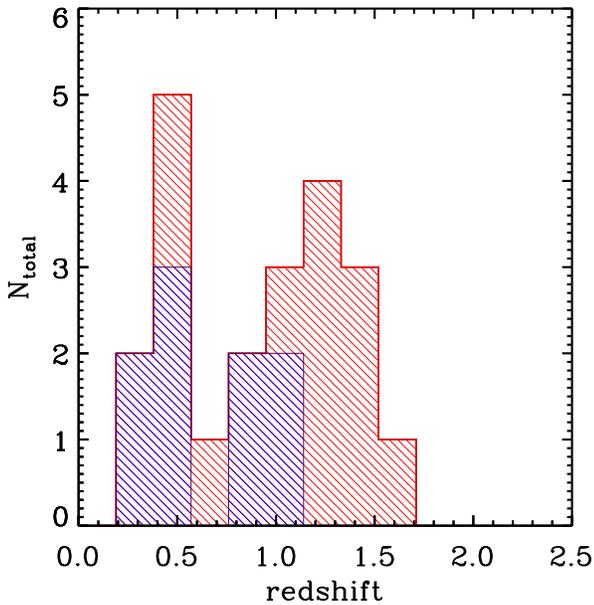}\\
\caption{Redshift distribution of sources in the MIPS dropout sample. A KMM test suggests a bimodal
distribution centred at $z$ $\sim$ 0.4 and $z$ $\sim$ 1.3. Blue shadowed area corresponds to the distribution of sources 
with spectroscopic redshift. }
\label{fig:sfsed}
\end{figure}\begin{figure}
\centering
\includegraphics[scale=0.36]{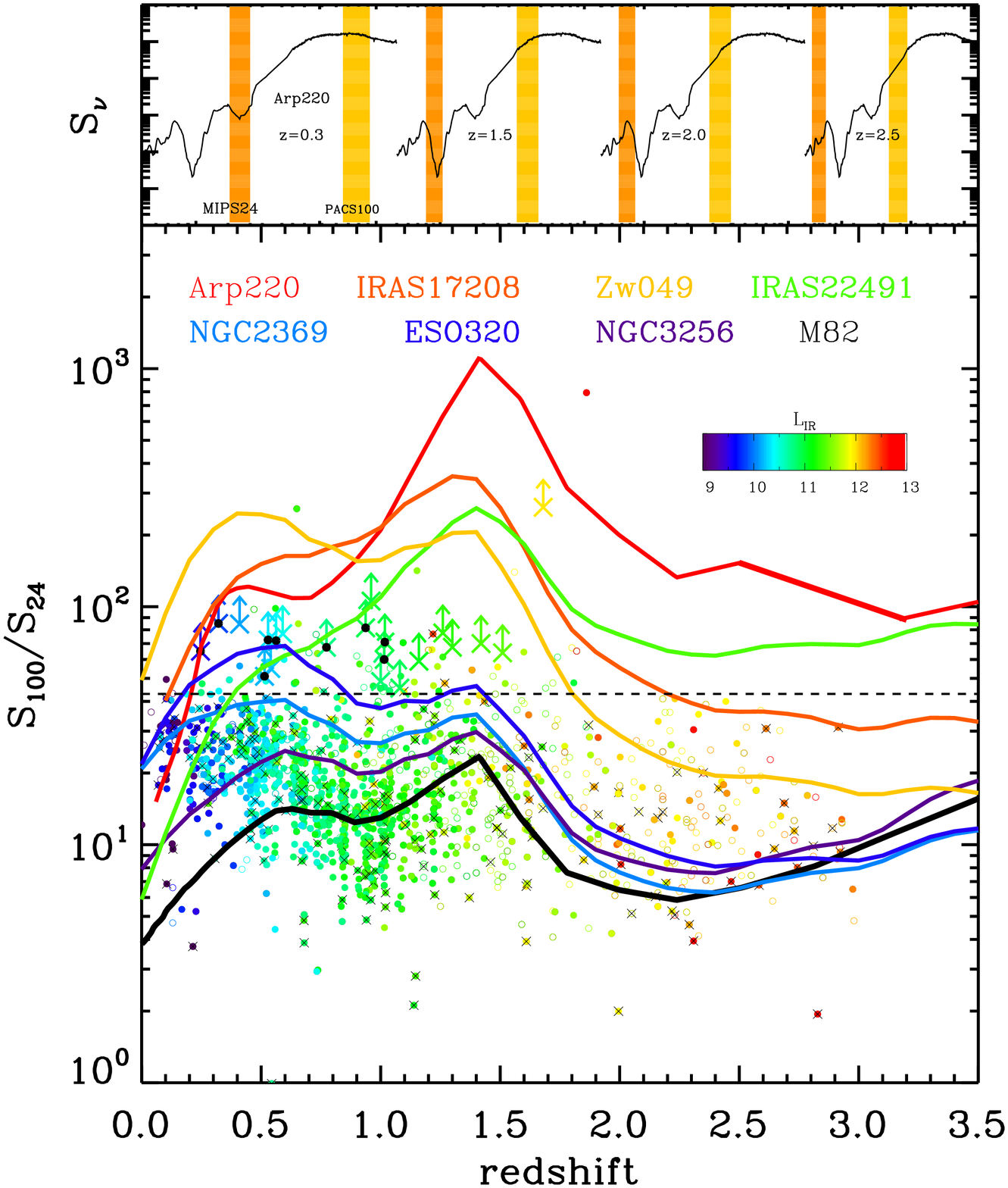}\\
\caption{\small{\pam ~as a function of redshift for the whole GOODS-H sample (circles) and lower limits for the $24\,\mu$m dropouts (arrows). Both samples are colour coded based on their \lir . Sources with a black cross are AGNs based on their X-ray emission. Filled symbols denote sources in the GOODS-H sample with spectroscopic redshift, while open symbols are sources with photometric redshift. Similarly, black circles on top of the arrows indicate that a spectroscopic redshift is available for that MIPS dropout source. Solid lines correspond to different observed SEDs of local LIRGs/ULIRGs (see Fig. 5) and horizontal black dashed line to \pam ~= 43. The pop panel shows the SED of Arp220 at various redshifts along with the MIPS 24- and PACS $100\,\mu$m bands.}}
\label{fig:sfsed}
\end{figure}
Similarly to Fig. 2, we now plot the \pam ~colour as a function of redshift (Fig. 4). In this figure we also overplot the \pam ~colour as a function of redshift for a number of local LIRGs/ULIRGs based on their observed SED as constructed by IRS observations of their mid-IR spectrum (Armus et al. 2007) and {\it IRAS} observations of their far-IR emission (Rieke et al. 2009). The observed templates were chosen to span a wide range of \lir ~and silicate optical depths ($\tau_{9.7}$) and their full SEDs are presented in Fig. 5, in order of increasing $\tau_{9.7}$. The $\tau_{9.7}$ measurements are adopted from Armus et al. (2007) (Arp200, NGC 22491), Pereira-Santaella et al. (2010) (ESO 320-G030, NGC 2369, Zw 049.057, NGC 3256) and da Cunha et al. (2010) (IRAS 17128). On the top of Fig. 4 we also show the SED of Arp220, along with the 24- and $100\,\mu$m bands at several redshifts. It is evident that  the \pam ~colour of the templates varies significantly as a function of redshift, mainly due to the presence of the silicate absorption features at 9.7- and $18\,\mu$m that enter the MIPS $24\,\mu$m filter at $z$ $\sim$ 1.4 and $\sim$ 0.3 respectively. We also note a wide range of \pam ~colours for a given redshift, indicative of different amounts of extinction as well as different dust temperatures. Indeed, Fig. 5 suggests  a weak trend for SEDs of sources with deeper silicate absorption to peak at shorter wavelengths. Therefore, the \pam ~values increase because \mips ~is suppressed by the silicate absorption features, but are also further elevated due to higher dust (big grain) temperatures  and hence \pacs ~values.

Looking back at Fig. 4, we see that $24\,\mu$m dropouts, although they have higher \pam ~values when compared to the whole sample, exhibit colours that are consistent with those of local star formation dominated ULIRGs/LIRGs. The fact that the redshift distribution of $24\,\mu$m dropouts peaks at redshifts where the 9.7- and $18\,\mu$m silicate absorption features enter the $24\,\mu$m band indicates that these sources could have moderate/strong silicate absorption features. On the other hand, the high \pam ~ratio could also result from lower levels of observed dust continuum emission at 24$\,\mu$m, but as we will see later that our data disfavour this scenario. In any case, all $24\,\mu$m dropouts fall within the envelope defined by the templates, so they form an extreme rather than an extraordinary population of star forming galaxies. We also note that none of the drop-out sources have an X-ray detection or meet the criteria for a power$-$law AGN.

Previous studies have attempted to identify infrared luminous sources that are undetected at $24\,\mu$m by combining 
$16\,\mu$m IRS peak-up imaging with $24\,\mu$m data (Kasliwal et al. 2005, Teplitz et al. 2011) and following an approach  similar to the one presented in this work. They searched for sources at $z$ $\sim$ 1.3, that are faint at $24\,\mu$m, due to the shift of the silicate absorption features into the MIPS band, but bright at $16\,\mu$m due to $7.7\,\mu$m PAH emission, concluding that sources with \irm ~$>$ 1.2 tend to be found at $z$ $>$1.1. They also investigated the \irm ~colours of several local LIRGs/ULIRGs and report an average ratio of $\sim$1-2 for galaxies with strong silicate absorption features. From our $24\,\mu$m dropout sample, none of the sources are detected at $16\,\mu$m down to the 3$\sigma$ detection limit (\irs ~$\sim$ 40 $\mu$Jy for GOODS-N and $\sim$ 65 $\mu$Jy for GOODS-S, Teplitz et al. 2011). Apart from the fact that some of our sources are outside the area covered by the IRS peak-up image in the GOODS fields, the non-detection of the rest of the sources  at $16\,\mu$m is somewhat expected from the discussion above. Even if we adopt a \mips  ~= 20 $\mu$Jy for all sources in the sample and assume a \irm ~= 2.0 (Arp220 case, Armus et al. 2007) then our sources should only be marginally detected at the depth of the $16\,\mu$m GOODS maps. 
\begin{figure}
\centering
\includegraphics[scale=0.42]{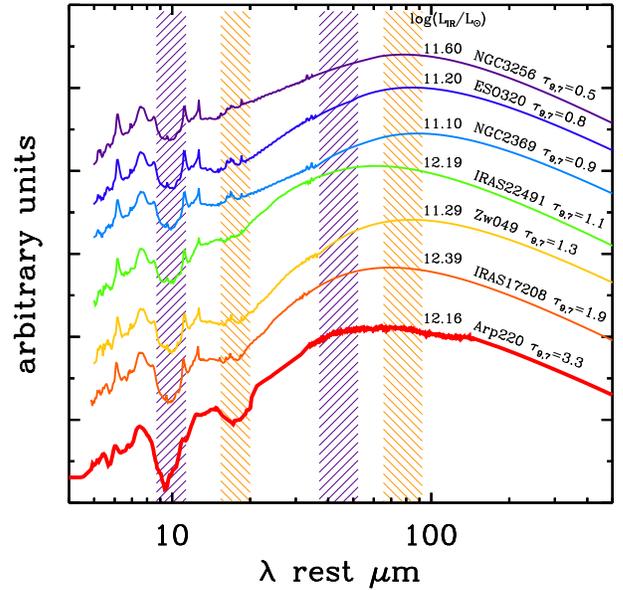}\\
\caption{A large range of rest frame SEDs of local ULIRGs and LIRGs, presented in increasing order of silicate optical depth. The mid-IR part is the IRS spectrum (Armus et al. 2007), while the far-IR comes from Rieke et al. (2009) templates. Blue (orange) stripes correspond to MIPS 24- and PACS $100\,\mu$m bands at $z$=1.3 (0.3)}
\label{fig:sfsed}
\end{figure}
\subsection{A $z$ $\sim$ 1.68 ULIRG, missed by MIPS} 
Although the study of the properties of individual $24\,\mu$m dropout sources  is beyond the scope of this study, here we wish to have a closer look at a specific source for which, \h data indicate an infrared luminosity \lir ~$>$ 10$^{12}$ \lsol. This is the only source in our high-$z$ sample with detection at $100\,\mu$m (\pacs ~= 5.2 $\pm$ 0.42 mJy), $160\,\mu$m ($S_{\rm 160}$ = 6.9 $\pm$ 1.1mJy) and at $250\,\mu$m. The source also has also a $\sim$ 6$\sigma$ detection at the GOODS-N VLA 1.4GHz map ($S_{\rm 1.4}$ = 32.1 $\pm$5.1 $\mu$Jy). 
Unfortunately, the source is out of the area covered by IRS $16\,\mu$m peak-up imaging. Cut out images of this galaxy  are shown in the sixth row of Fig. 1. To derive the far-IR properties of the source we fit the observed \h ~points with CE01 models, but also with a range of observed SEDs of local ULIRGs, described in the previous section. The observed multi-band photometry, together with the best fit SEDs 
are shown in Fig. 6.

The derived infrared luminosity of the source is \lir ~= 1.3$-$1.6 $\times$ 10$^{12}$ \lsol ~(depending on the assumed SED), but it is evident that only templates with a strong silicate absorption feature at $9.7\,\mu$m can reproduce the non-detection down to 20 $\mu$Jy at $24\,\mu$m. Fitting the optical part of the SED with BC03 models, yields a stellar mass of 8.9 ($\pm$0.5) $\times$ 10$^{10}$ \msol, ~while based on the $\beta$ slope of the UV spectrum we derive a reddening {\it E(B$-$V)} = 2.1, consistent with a heavily obscured source. Assuming a Salpeter IMF and the Kennicutt (1998) relation we convert the \h $-$based \lir ~to star formation rate and derive  SFR $\sim$ 280 M$_{\odot}$ yr$^{-1}$. Similarly, we convert the radio flux to \lir ~(Condon 1992) and subsequently to SFR, finding that the two estimates are in perfect agreement ($SFR_{\rm radio}$ = 300 $\pm$ 32 \msol ~yr$^{-1}$). The source is not detected in X-rays and exhibits the $1.6\,\mu$m stellar bump, indicating that there is no near-to-mid-IR evidence for the presence  of an AGN. The above analysis suggests that this is a heavily obscured starburst galaxy with no signs of AGN activity. Although unique in our sample, this galaxy raises the question of how many such objects we might have missed in the pre-\h ~era and what would be their contribution to the cosmic SFR density. A detailed discussion on this point will be presented in section 5.   

\begin{figure}
\centering
\includegraphics[scale=0.42]{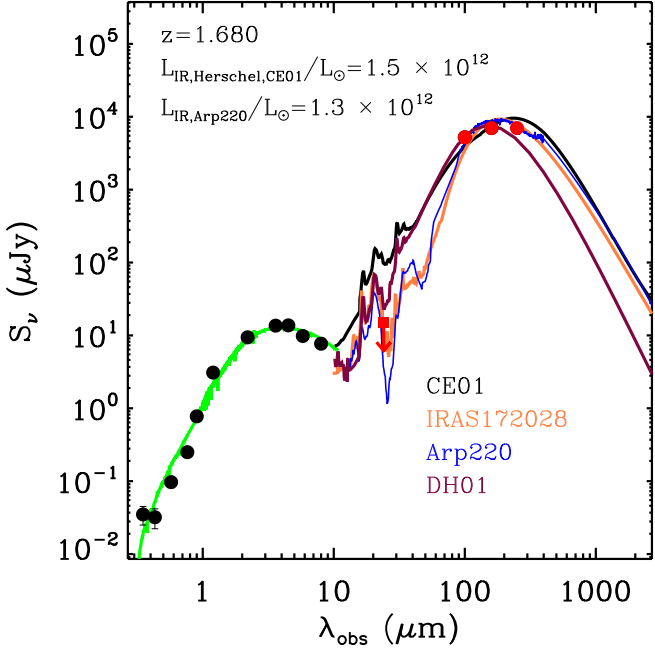}\\
\caption{\small{The SED of a dropout at $z=1.68$, detected in both the 100- and $160\,\mu$m bands, although missed at $24\,\mu$m. The optical part is overlaid with the best fit BC03 model (green line) and the infrared part with a range of observed and model SEDs. The red square denotes the upper limit at $24\,\mu$m. Cut-out images of the source are shown in Fig. 1, 6th row.}}
\label{fig:sfsed}
\end{figure}

\section{{\it Silicate-break} galaxies}
So far, we have demonstrated that \h ~data have revealed a small but interesting
 class of infrared luminous galaxies, which are undetected at $24\,\mu$m. The main characteristic of these sources is their atypically red  \pam ~colour. Furthermore, Fig. 2 (left), reveals that sources with \pam ~colours similar to those of the dropouts can be found among sources with $24\,\mu$m detection. In what follows we extend our study to the whole GOODS-H sample, searching for such objects. 
   
To select them, we adopt a cut off, \pam ~$>$ 43, that corresponds to the bluest lower limit among the MIPS dropout sample and the colour of the local LIRG ESO320 shifted at $z=1.4$. In Fig. 7 we plot the redshift distribution of sources with \pam ~$>$ 43 (excluding the dropouts), finding that it bears a remarkable resemblance to that of the $24\,\mu$m dropouts in Fig. 3. Indeed, a K-S test reveals that there is no significant difference between the two samples with a $p$ value  of 0.61. In contrast, we find that the sample has a redshift distribution quite different than that of the whole  GOODS-N sample (Fig. 7, top panel) at a confidence level of $>$97.5$\%$. Similar to the dropout sample, sources with \pam ~$>$ 43 exhibit a bimodal redshift distribution at the 3.5$\sigma$ level, centred at $z$ $\sim$ 0.48  and $\sim$ 1.3. Therefore, in what follows, 
we suggest that the \pam ~colour could serve as redshift indicator.
\begin{figure}
\centering
\includegraphics[scale=0.43]{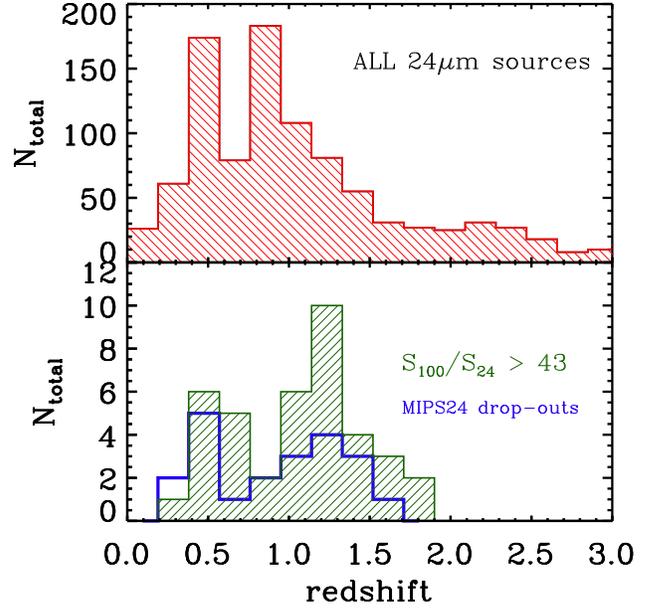}\\
\caption{Top: Redshift distribution of sources in GOODS-N with PACS $100\,\mu$m and MIPS $24\,\mu$m detections.
Bottom: Redshift distribution of sources in GOODS-N with \pam ~$>$ 43. Sources are clustered around $z$ = 0.4 and $z$ =1 .3. The blue histogram corresponds to the redshift distribution of MIPS dropouts. }
\label{fig:sfsed}
\end{figure}

\subsection{\pam ~colour as a redshift indicator.}
In the pre-\h ~era, there were several attempts to use anomalous 
MIR colours as a crude redshift indicator.
These studies mainly employed the $16\,\mu$m IRS peak-up imaging and proposed that a  blue \irm ~colour would peak in a narrow redshift bin, 1.0 $<$ $z$ $<$ 1.8  (Armus et al. 2007, Teplitz et al. 2011), as well as being useful for selecting infrared luminous galaxies in this redshift bin that were undetected by the $24\,\mu$m band. The main idea behind this criterion is that at these redshifts the $9.7\,\mu$m silicate absorption feature enters in the $24\,\mu$m passband, while the strong PAH emission features at 6.2- and 7.7 $\mu$m 
are shifted into the $16\,\mu$m band, producing a distinctive blue \irm ~colour.  
For example, Kasliwal et al. (2005) used a  \irm ~$>$ 1.2 ratio to select objects 
at 1.1 $<$ $z$ $<$ 1.6, and called them ``silicate-break galaxies'', attributing their blue \irm ~colour to the existence of a strong silicate absorption feature at $9.7\,\mu$m. This approach had two main caveats. First, the blue \irm ~colour is not necessarily produced by a silicate absorption feature, as similar colours can 
appear for sources with strong PAHs and low dust continuum. Second, many objects at lower redshifts fall within the same colour cut (Teplitz et al. 2005).  A more 
recent attempt by Teplitz et al. (2011), reports that  a higher ratio (\irm ~$>$ 1.4), would eliminate many but not all of the low-$z$ interlopers.

Here, we face the same situation. The cut in the \pam ~ratio that we have adopted selects sources in two redshifts bins. In order to reject the low-$z$ sources in our sample, simply increasing the \pam ~cut is not useful as apart from the low-$z$, we also miss many high-$z$ sources and the selection is still not pure enough. Alternatively, we can employ a second colour criterion, based on the $16\,\mu$m and $8\,\mu$m flux densities. In Fig. 8  we plot the \pam ~vs \iri ~colour $-$ colour diagram for sources 
that have a 16- and $8\,\mu$m detection. We see that if a ratio cut of \iri $>$ 4 is combined with a \pam ~cut $>$ 43, then we successfully reject all 
low-$z$ interlopers, while selecting sources in the 1.0 $<$ $z$ $<$ 1.7 redshift bin. The selection of  \iri ~as a second colour criterion was driven 
by the fact that at 1.0 $<$ $z$ $<$ 1.7 the $16\,\mu$m band probes the PAH complex (6.2 $-$ 7.7$\,\mu$m), boosting the value of \irs ~while for the low redshift sources only traces 
emission from a warm dust continuum at $>$ $10\,\mu$m.  We note that none of our sources is classified as an AGN, based on their X-ray emission, 
their optical spectra or their mid-IR colours (i.e. power law AGNs). It therefore seems that we have found a way to select  star-forming
high-$z$ galaxies in a narrow redshift bin. In what follows we argue that these sources are compact starbursts with moderate/strong silicate features in their MIR spectrum.
 
\begin{figure}
\centering
\includegraphics[scale=0.4]{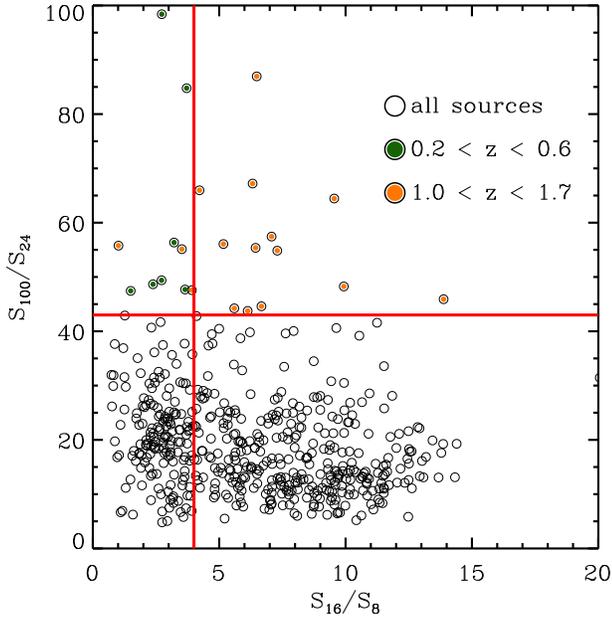}\\
\caption{\pam ~vs \iri ~colour $-$ colour diagram for the whole GOODS-H sample in GOODS-N with $16\,\mu$m detection (black circles). Circles filled with green and orange colour are sources with 
\pam ~$>$ 43 at 0.2 $<$ $z$ $<$ 0.6 and 1.0 $<$ $z$ $<$ 1.7, respectively. 
Vertical and horizontal red lines indicate a cut off ratio of \iri ~$>$ 4 and \pam ~$>$ 43. We see that this diagram can serve as a redshift diagnostic, for sources at 1.0 $<$ $z$ $<$ 1.7. }
\label{fig:sfsed}
\end{figure}
\begin{figure}
\centering
\includegraphics[scale=0.4]{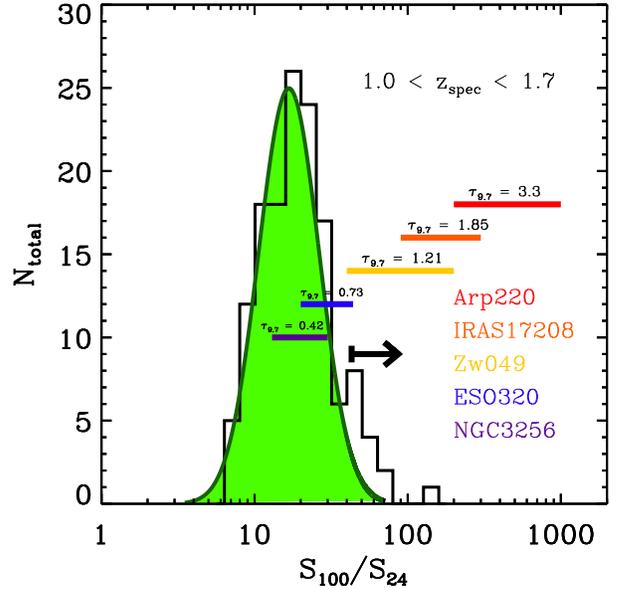}
\caption{\small{Distribution of the \pam ~colour among sources in the GOODS-H sample with spectroscopic redshift at 1.0 $<$ $z$ $<$ 1.7. 
Green line and green shadowed area indicates the Gaussian fit to the distribution and the area beneath it. Coloured bars indicate the range of the \pam ~colour of several observed SEDs of local LIRGs and ULIRGs at this redshift bin. The silicate optical depths of the local sources, as derived by IRS spectroscopy are also overlaid. The black arrow indicates the position of the high-$z$ $24\,\mu$m dropout sample.}}
\label{fig:sfsed}
\end{figure}   
\subsection{Evidence for silicate absorption}

We have already discussed in the Introduction,  that for normal galaxies, i.e. those with \lir ~$<$10$^{10}$ \lsol,  discriminating between 
moderate PAH emission superposed on a silicate-absorbed continuum and strong PAH features with 
a relatively weak underlying continuum is a difficult task (e.g. Smith et al. 2007), even when high quality mid-IR spectra are available. 
A typical example is M82, for which Sturm et al. (2000)  suggests that there is no silicate absorption
since the $10\,\mu$m dip can be reproduced by a superposition of strong PAHs and VSG continuum.
On the other hand, things are more straightforward for more luminous infrared  sources, where there is clear evidence for the existence of a wide range 
of silicate optical depths, both for local and high-$z$ LIRGs and ULIRGs 
(Armus et al. 2007, Pereira-Santaella et al.
2010).  Unlike the sample based on the blue \irm ~colour, for which the \lir ~estimates were based on large/uncertain extrapolations, our study benefits from more robust \h$-$based  \lir ~estimates. The fact that  all sources with \pam ~$>$ 43  at 1.0 $<$ $z$ $<$ 1.7  
in our sample (including the high-$z$ $24\,\mu$m dropouts) have \lir ~$>$ 10$^{11}$ \lsol, coupled with the lack of observational evidence 
of sources with similar luminosities and low dust continuum emission, is a first hint that these sources have a silicate absorption feature at $9.7\,\mu$m.

We have already demonstrated in Fig. 4, that these sources have \pam ~colours consistent with those of  
local templates of LIRGs and ULIRGs with moderate/high silicate strength. To further investigate this, we plot the distribution of the \pam ~colour of the GOODS-H 
(GOODS-N and GOODS-S) sources with spectroscopic redshift at 1.0 $<$ $z$ $<$ 1.7, along with the range of the \pam ~colour at this redshift range
for 5 LIRG/ULIRG observed templates with a wide range of $\tau_{9.7}$ values (Fig. 9). 
In practice, this plot examines the detailed distribution of galaxies in the 1.0 $<$ $z$ $<$ 1.7 region of Fig. 4. 
Fitting the colour distribution with a Gaussian indicates a clear excess in the red tail due to sources with \pam ~$>$ 43. 
It is crucial to stress that this excess is absent  in other redshifts bins. Furthermore, we see that based on the template SEDs, 
sources with higher optical depth exhibit redder \pam ~colours in this redshift bin. In other words, it seems that the \pam ~colour could serve 
as a rough indicator of the silicate strength. Our sources, i.e. those with \pam ~$>$ 43, have colours consistent with 0.7 $<$ $\tau_{9.7}$ $<$ 1.9, 
while it appears that none of them have a silicate feature as strong as that of Arp220. We note however, that we cannot rule out a source in our sample with 
stronger silicate but colder $T_{\rm dust}$. 

An alternative explanation of the high \pam ~ratios of these sources would be enhanced $100\,\mu$m emission, simply because the sources are warmer, and hence their SEDs peak at shorter wavelengths. To test this, we investigate the \pam ~colour as a function of dust temperature, as indicated by the \temp ~ratio.  In Fig. 10,  we present this colour$-$colour diagram 
for sources at  1.0 $<$ $z$ $<$ 1.7, which are detected at 160-and $100\,\mu$m. It appears that the adopted \pam ~cut does not introduce a bias towards 
warmer sources (i.e. sources with low \temp ~colours), as it selects sources with a wide range of dust temperatures similar to that found 
for the whole GOODS-H sample in this redshift bin, suggesting that an enhanced \pacs ~is not the main reason for their high \pam ~ratios. We note that 
a large dispersion of dust temperature of high-$z$ galaxies has recently been demonstrated by Hwang et al. (2010) and Magdis et al. (2010d).
\begin{figure}
\centering
\includegraphics[scale=0.41]{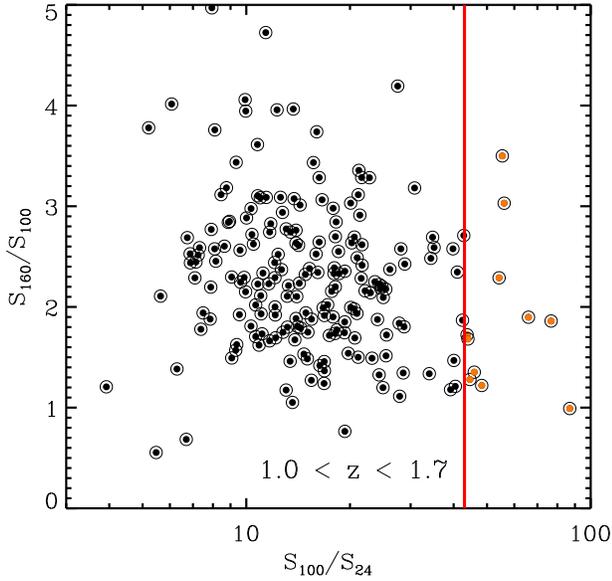}
\caption{\small{\temp ~vs \pam ~colour $-$ colour diagram for sources in the GOODS-H sample at 1.0 $<$ $z$ $<$ 1.7. Sources with \pam ~$>$ 43 and \iri ~$>$ 4 are filled with orange colour. 
This plot demonstrates that our selection is not biased towards warmer sources.}}
\label{fig:sfsed}
\end{figure}

One of the sources that meets the selection criteria is the well studied sub-millimetre source at $z$ = 1.21, GN26 for which Pope et al. (2008) have published an IRS spectrum. The detection of this source in the PACS bands has also been discussed by  Dannerbauer et al. (2010) and Magnelli et al. (2010) while Frayer et al. (2008) has reported the detection of CO(2$\rightarrow$1) emission. In Fig. 11, we show the full SED of the source, overlaid with the best CE01 template 
for the far-IR part and the observed IRS spectrum for the mid-IR part of the SED. 
The IRS spectrum of GN26 is also presented separately in an inset panel. This source has \pam ~$\sim$ 70, and according to Fig. 9, it should have 
a silicate optical depth of $\sim$ 1. Although the S/N of the IRS spectrum does not allow for a robust measurement of $\tau_{9.7}$, its IRS spectrum is 
very similar to that of the composite spectrum of SMGs presented in Pope et al. (2008), and for which they report  
$\tau_{9.7}$ $\sim$ 1 in excellent agreement with what we expected based on this source's \pam ~ colour. We should also note that GN26 was one of the 13 sources used  for the construction of the composite spectrum. Taken together, the evidence suggests that the red \pam ~colour of our sample  is caused by 
the existence of a moderate/strong silicate absorption feature at $9.7\,\mu$m, that enters the $24\,\mu$m band at these redshifts. 
We therefore choose to characterize these sources as  {\it silicate-absorbed} galaxies.
   \begin{figure}
\centering
\includegraphics[scale=0.4]{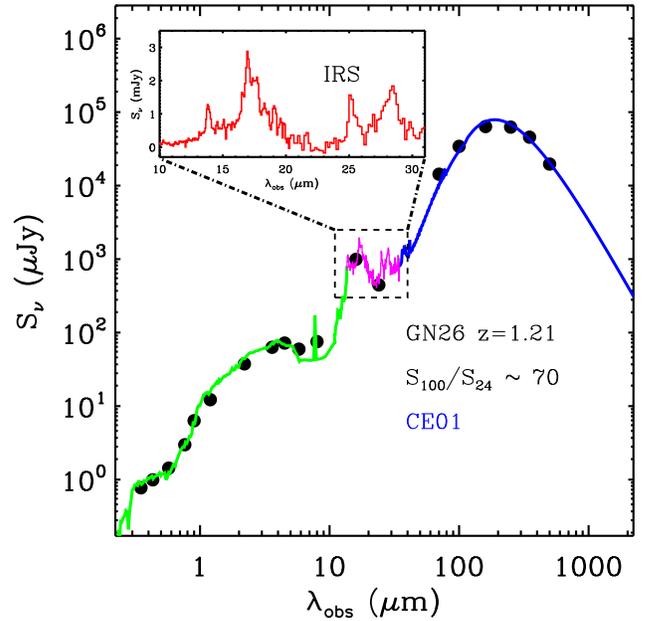}
\caption{\small{Full SED of GN26, a source at $z$ = 1.21  with \pam ~$\sim$ 70. The far-IR part of the SED is derived from \h ~data and is overlaid with the best fit CE01 model. The mid-IR part of the SED is the observed IRS spectrum from Pope et al. (2008), that is also presented separately in the inset figure. We also show the optical/near-IR part of the SED along with the best fit BC03 model.}}
\label{fig:sfsed}
\end{figure}
\begin{figure}
\centering
\includegraphics[scale=0.36]{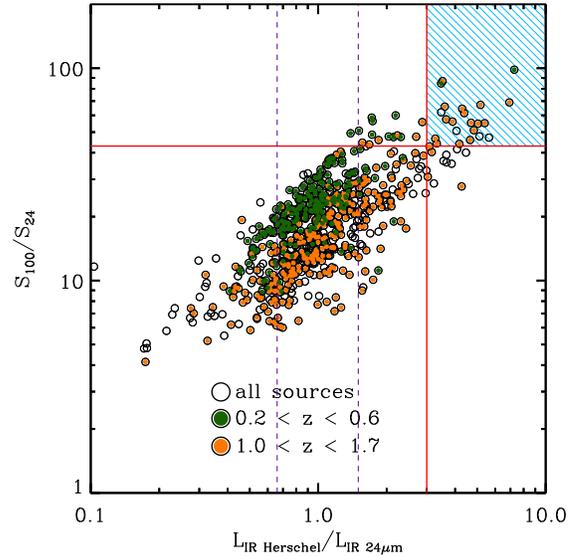}
\caption{$S_{\rm 100}$/$S_{\rm 24}$ as a function of the \h ~derived \lir ~over the \lir ~derived based only on the $24\,\mu$m flux densities. Green circles are sources at 0.2 $<$ $z$ $<$ 0.6 and orange circles are sources at 1.0 $<$ $z$ $<$ 1.7. The top right box  (defined by \pam ~$>$ 43 and $L_{\rm IR~Herschel}$ / $L_{\rm IR~24}$ $>$ 3 ) consists mainly of sources at 1.0 $<$ $z$ $<$ 1.7. Vertical dashed lines indicate the area where the two \lir ~estimates agree within a factor of 1.5.}
\label{fig:sfsed}
\end{figure}

\begin{figure*}
\centering
\includegraphics[scale=0.4]{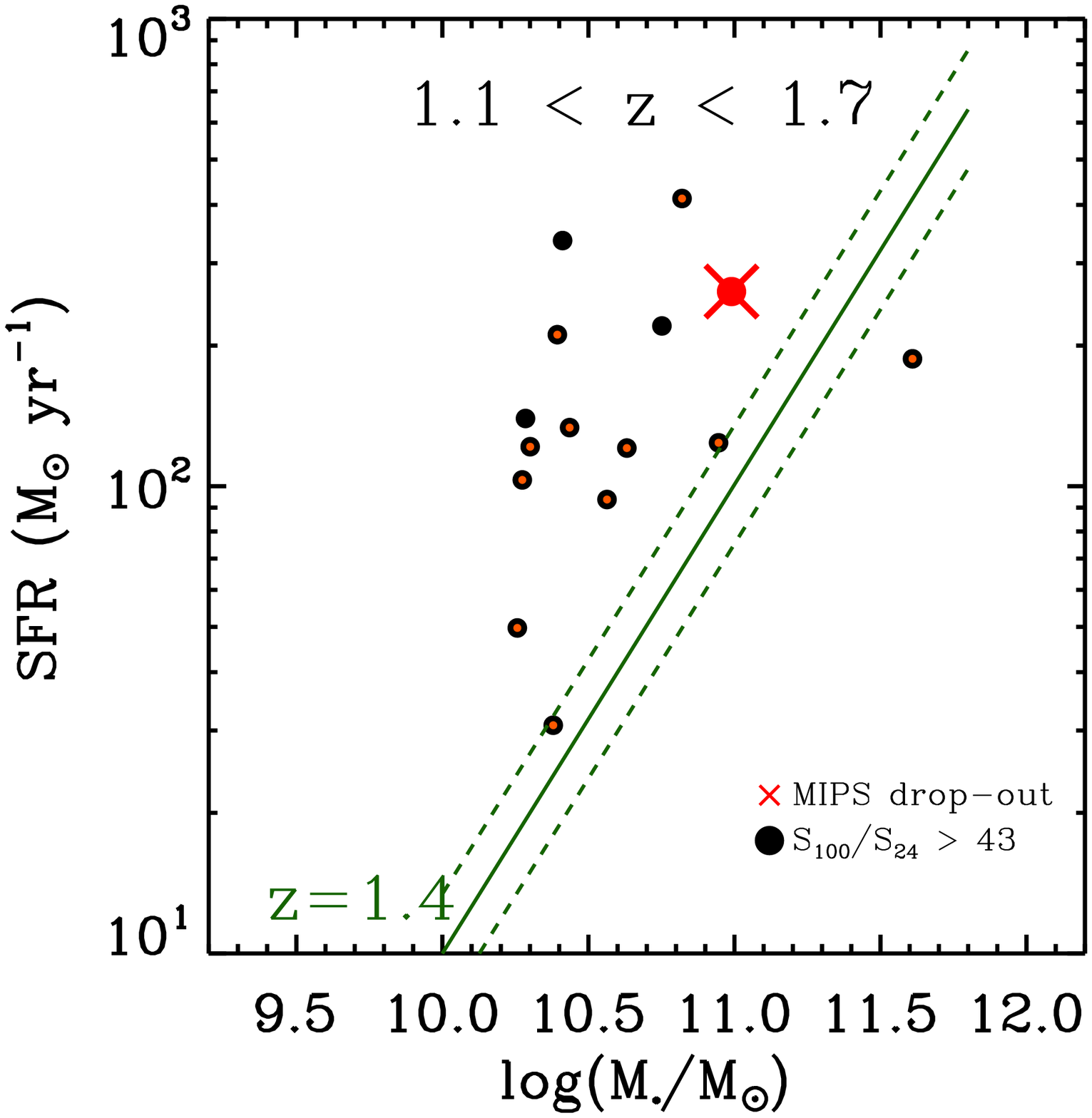}
\includegraphics[scale=0.4]{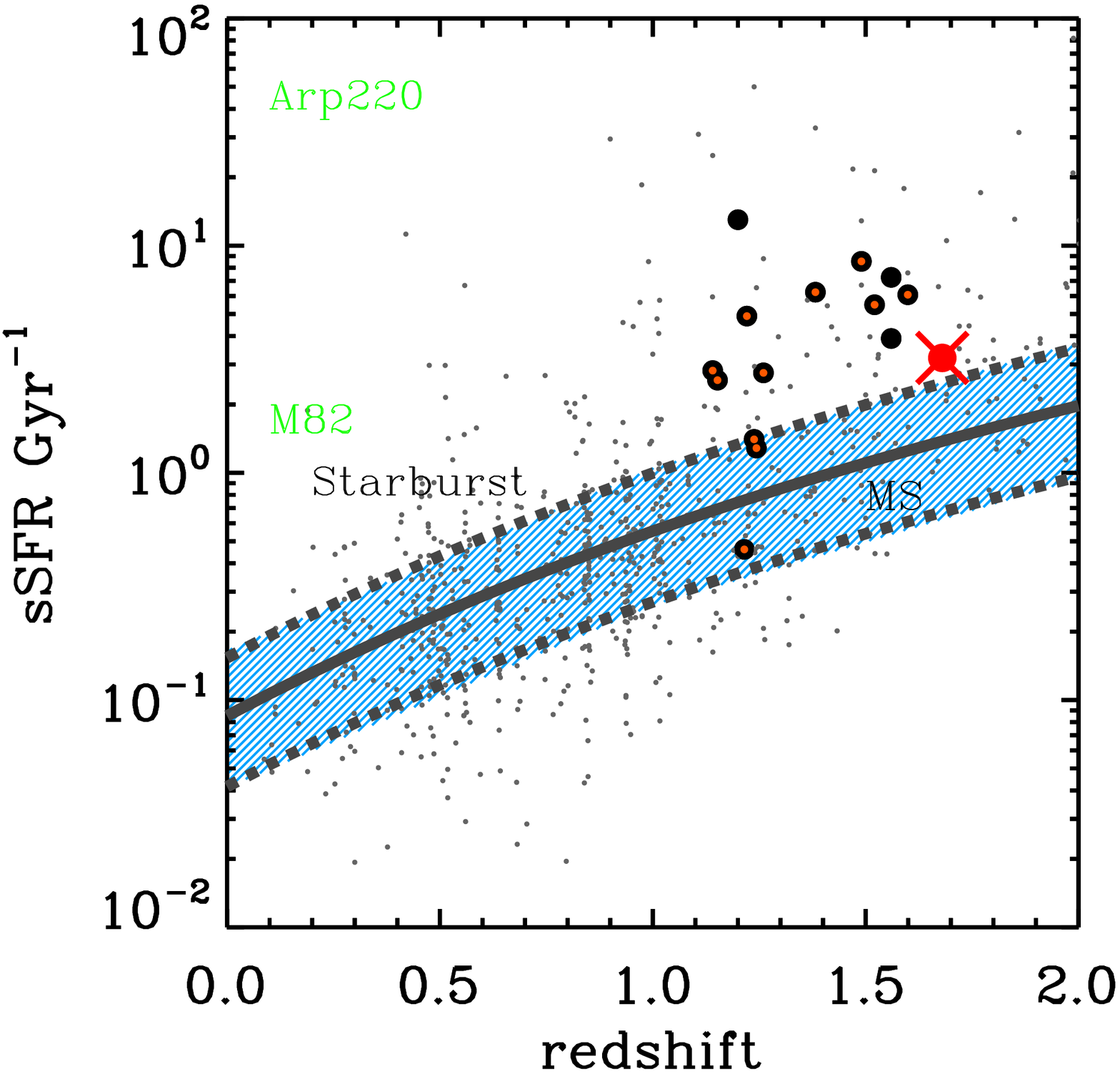}
\caption{Left: The SFR $-$ M$_{\ast}$ relation for sources with \pam ~$>$ 43 at 1.1 $<$ $z$ $<$ 1.7 (black circles). Orange dots indicate sources  with spectroscopic redshift. Red cross corresponds to the ULIRG dropout that we presented in Section 3.3. The solid green line depicts the SFR$-$M$_{\ast}$ correlation at $z$ $\sim$ 1.4 (Elbaz et al. 2011), while the dashed lines indicate its dispersion. Right: Specific star formation rate (sSFR), as a function of redshift, for the same sample as in the left panel, as well as for the whole GOODS-H sample (grey dots). The gray thick line denotes the evolution of the sSFR with redshift, as derived by individual detections and stacking analysis of the GOODS-H sample by Elbaz et al. (2011). The blue shaded area along with the dashed gray lines which correspond to the dispersion of the evolution ($\sim$ 0.3 dex), indicate the region of main sequence galaxies. Sources with  \pam ~$>$ 43 tend to be found above the blue shaded area, populating the starburst region. The green text indicates the position of Arp 220 and M82 (shifted to $z = 0.1$ for clarity) }
\label{fig:sfsed}
\end{figure*}
\subsection{The far-IR properties of the high-$z$ sample}
In the pre-\h ~era the far-IR properties of large samples of galaxies were derived based on large 
extrapolations of their $24\,\mu$m flux density using SED templates based on local IR-luminous galaxies. Recent studies using \h ~data have confirmed 
the validity of these extrapolations, demonstrating that the $24\,\mu$m flux density is a good proxy of the total \lir, at least up to $z$ $\sim$ 1.5 (Elbaz et al. 2010). On the other hand, here we have found a population of 
high-$z$ sources (1.0 $<$ $z$ $<$ 1.7), that exhibit atypically red \pam ~colours. For these sources, 
we expect that \lir ~estimates based on their \mips, and using average template SEDs, such as CE01 and DH02, would be severely underestimated.

In Fig. 12  we plot the \pam ~colour of all GOODS-H galaxies as a function of the ratio between the \h ~based \lir ~and the one derived using only the $24\,\mu$m flux density. Although for the majority of the sources the two \lir ~estimates are in good agreement 
(within a factor of $\sim$ 1.5), for the high-$z$ sources with \pam ~$>$ 43 the $24\,\mu$m flux density would underestimate the true \lir ~on average by a factor of $>$ 3. This is not the case for the 
low-$z$ sample where the two estimates are in better agreement. For the high-$z$ sample there is a clear 
trend between the \pam ~colour and the ratio of the two \lir ~estimates, in the sense that the true \lir ~is progressively underestimated for sources with redder \pam ~colours, depicting the limited variety of SEDs used for the derivation of the \lir. This indicates that 
 although the average template SEDs are representative for the bulk of the galaxy population at $z$ $<$ 1.5, this is not the case for a population at 1.0 $<$ $z$ $<$ 1.7 with very red \pam ~colours  ($>$ 43 for this study).

 \subsection{The starburst nature of the high-$z$ sample}
As the silicate absorption feature merely requires a mass of warm dust obscured by a significant column of cooler dust, it does not provide any insight into the mechanism that is  heating the warm dust. Hence, it is difficult to establish a correlation between the strength of this feature and the source that powers the mid-IR emission of infrared luminous galaxies, as it could equally be produced by a deeply buried AGN or a compact starburst (Farrah et al. 2008, Imanishi et al. 2009, Armus et al. 2007).  

We have already reported than none of our sources show direct signs of AGN activity, since none of them are either (a) detected in X-rays in the 2 Msec Chandra observations (Alexander et al. 2003), (b) satisfies the criteria of power$-$law AGN or (c) has high excitation lines in their optical spectra (where available). A deeply obscured AGN though, cannot be ruled out. To investigate this, we stack the Chandra X-ray data on these sources. We find a strong detection (7$\sigma$) in the 0.5$-$2 keV band corresponding to $L_{\rm 2-10}$ = 3 $\times$ 10$^{41}$ erg sec$^{-1}$ and no detection in the 2$-$7keV band (3$\sigma$ upper limit 1.28 $\times$10$^{-17}$ erg cm$^{-2}$ s$^{-1}$). The derived upper limit in the hard band along the stacking results in the soft band suggest that the sources are dominated by star formation (Nandra et al. 2002, Lehmer et al. 2008). Furthermore, in most cases the radio based \lir ~is in good agreement with that derived by \h, again indicating that those sources are dominated by a nuclear starburst, rather than an AGN. 

It has recently been shown that normal star$-$forming galaxies exhibit a correlation between their
SFR and stellar mass at any given redshift. This correlation was first found among $z$ $\sim$0 galaxies (Brinchmann et al. 2004) and was subsequently extended to higher redshifts, $z$ $\sim$ 1 Elbaz et al. (2007), Noeske et al. (2007), $z$ $\sim$ 2 Daddi et al. (2007), Pannella et al. (2009), $z$ $\sim$ 3 Magdis et al. (2010a,b,c) and  $z$ $\sim$ 4 Daddi et al. (2009). It has also been shown that star forming galaxies that do not follow this correlation tend to undergo a rapid starburst phase and have more compact geometries (Elbaz et al. 2011). For instance, at $z$ $\sim$ 0 and z $\sim$ 2, respectively, local ULIRGs and SMGs have SFRs that greatly exceed the SFR$-$M$_{\ast}$ correlation. Both classes of objects are thought to host compact starbursts (Daddi et al. 2010, Tacconi et al. 2010). It would therefore be interesting to investigate the position of the sources with red \pam ~colour in the SFR - M$_{\star}$ diagram. 
\begin{figure*}
\centering
\includegraphics[scale=0.39]{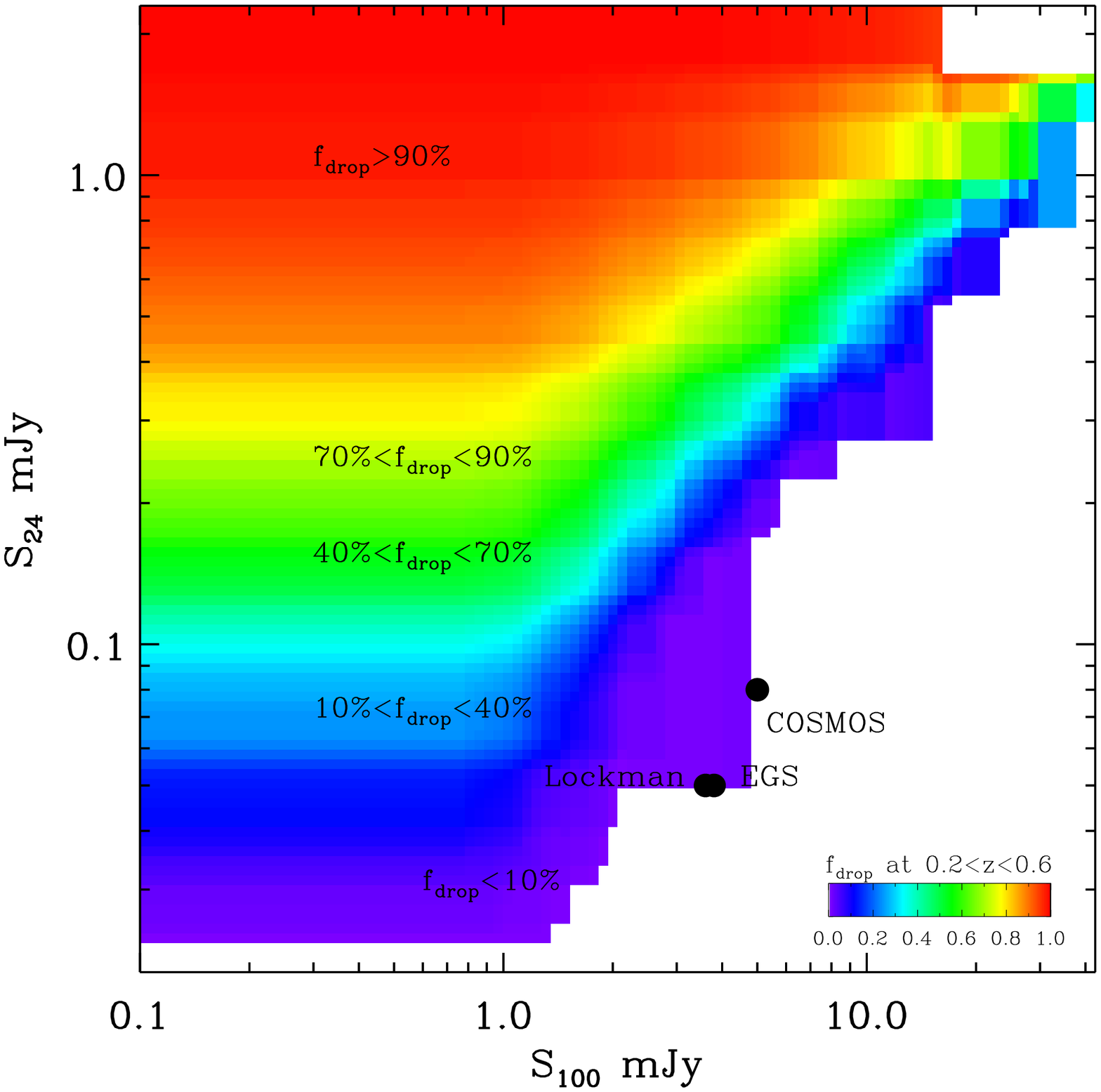}
\includegraphics[scale=0.39]{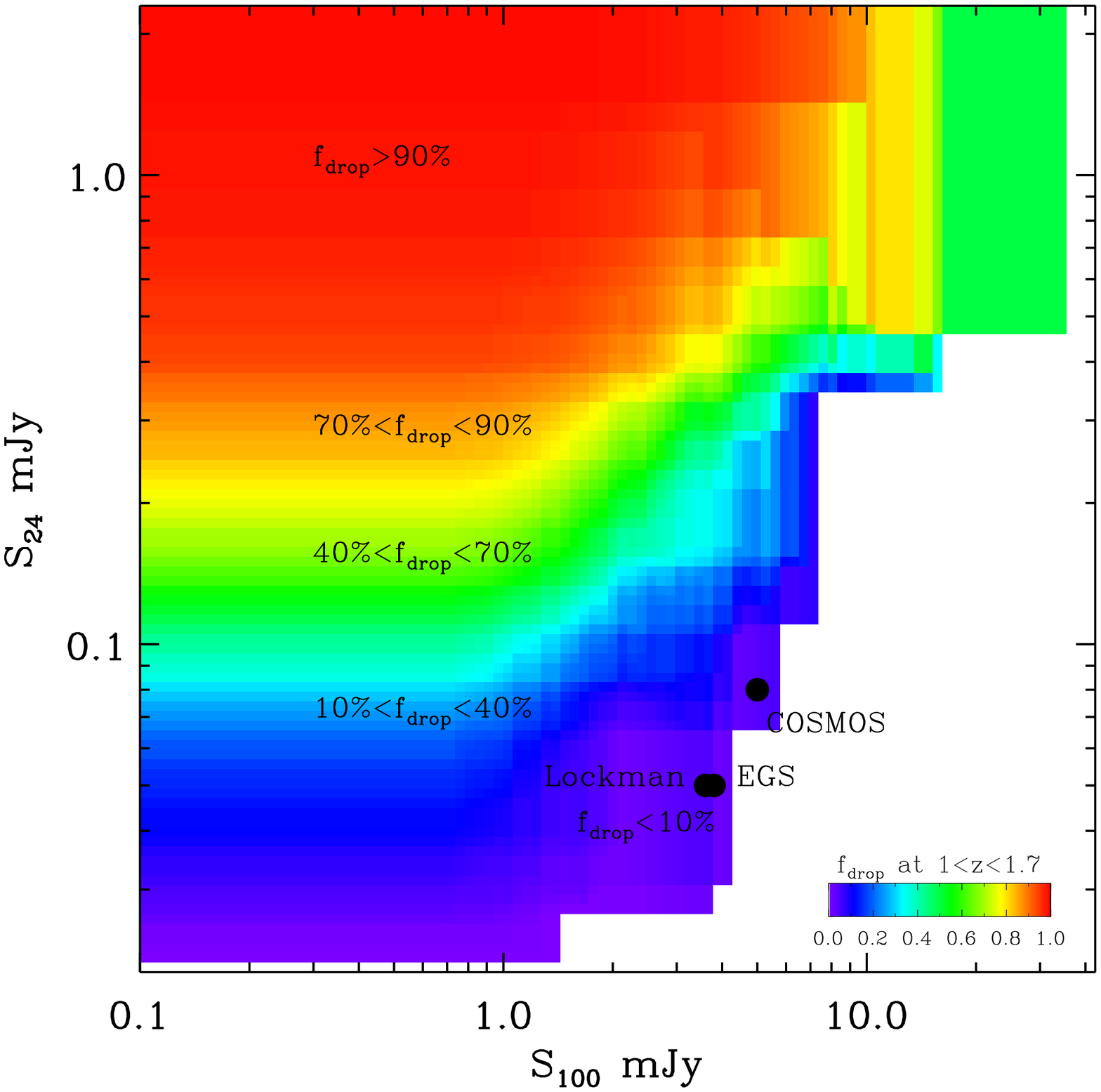}
\includegraphics[scale=0.39]{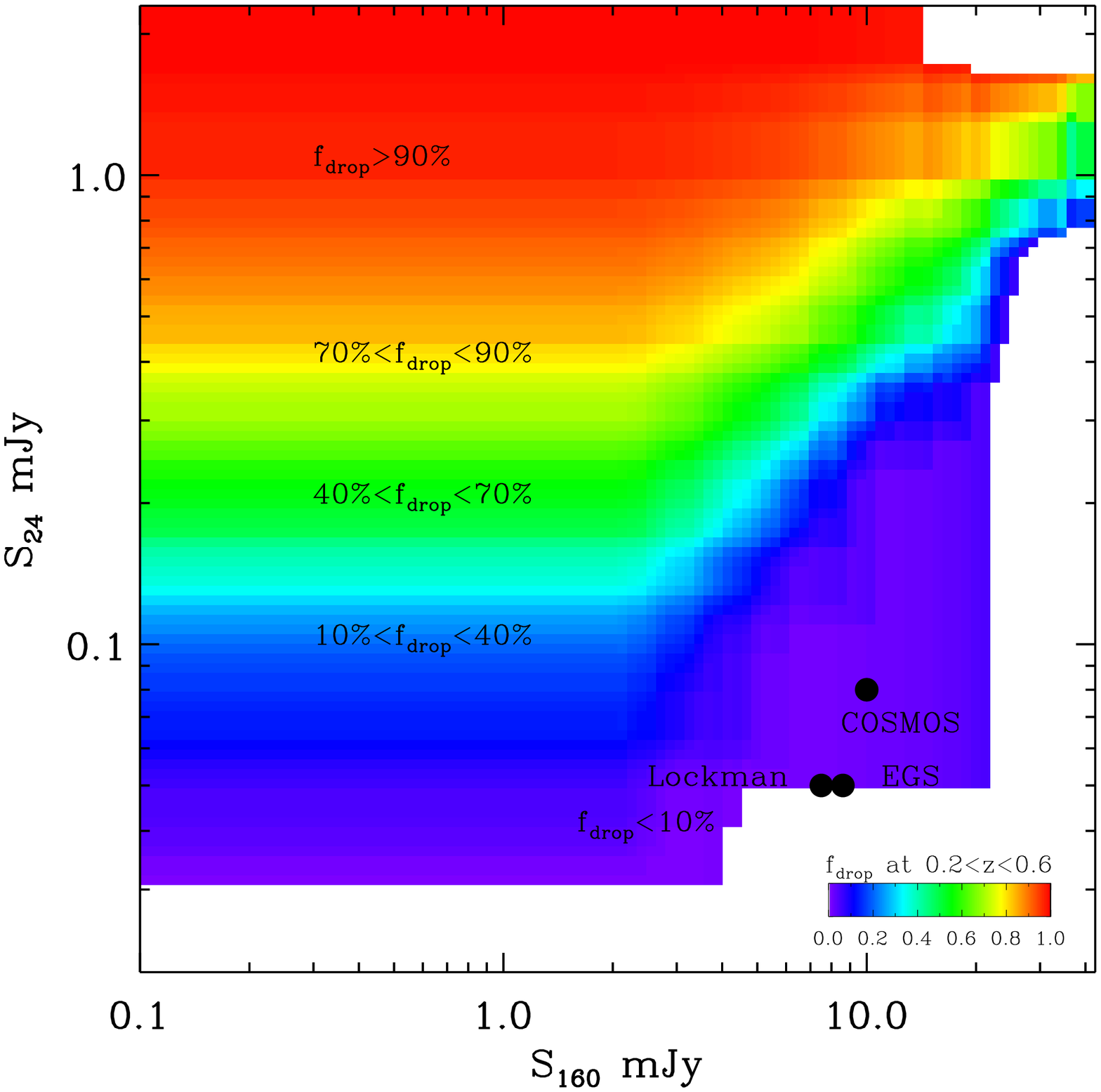}
\includegraphics[scale=0.39]{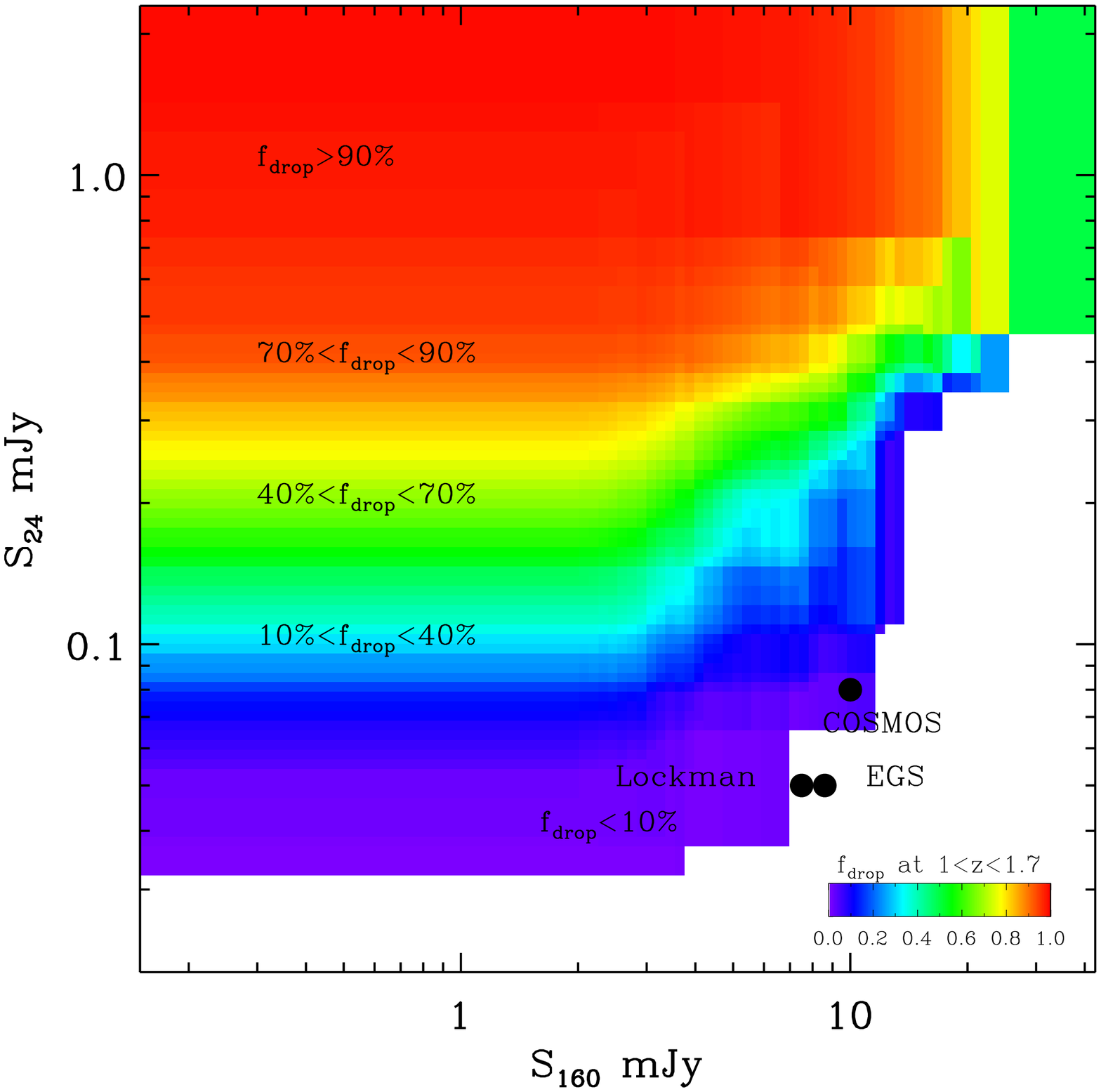}
\caption{The fraction of sources expected to be missed in the $24\,\mu$m band at 0.2 $<$ $z$ $<$ 0.6 (left) and 1.0 $<$ $z$ $<$ 1.7 (right)  as a function of the $24\,\mu$m vs $100\,\mu$m (top) and the $24\,\mu$m vs $160\,\mu$m (bottom) depths. Black circles indicate the depths of PACS and MIPS$24\,\mu$m surveys for some of the most important cosmological fields, i.e EGS, Lockman and COSMOS. We note that the region where the constant fraction isocontours run horizontally, cannot be constraint by the data. The adopted detection limits (5 $\sigma$) at 24$\,\mu$m are 80$\,\mu$Jy for COSMOS and 50$\,\mu$Jy for EGS and Lockman. Similarly the 3 $\sigma$ detection limits at 100$\,\mu$m are 5.0-, 3.8- and 3.6mJy (COSMOS, EGS and Lockman), while at 160$\,\mu$m are 10.2-, 8.6- and 7.5mJy.  }
\label{FigGam}%
\end{figure*}
Using the Kennicutt 1998 relation and a Salpeter IMF, we convert the \h ~based \lir ~to SFR for the sources with \pam ~$>$ 43 and 1.1 $<$ $z$ $<$ 1.7 and we plot the derived SFR versus the stellar mass of the galaxies (Fig. 13, left). We also overplot the SFR$-$M$_{\star}$ correlation at the median redshift of the sample i.e., $z$ = 1.4 (Elbaz et al. 2011). Clearly the sources are off the correlation, exhibiting enhanced star formation activity for their stellar mass. Similarly the ULIRG dropout source that we presented in section 3.3 is also off the correlation. To eliminate the effect of the evolution of the SFR$-$M$_{\star}$ with redshift, 
we also show the specific star formation rate (sSFR), defined as SFR/M$_{\ast}$, as function of redshift for sources with \pam ~$>$ 43, as well as for the whole GOODS-H sample (Fig. 13, right). We also overplot the evolution of sSFR with time as derived by Elbaz et al. (2011), based both on detected sources and stacking analysis of the GOODS-H sample. According to Elbaz et al. (2011), sources within the blue shaded area are main sequence, normal star forming galaxies galaxies while sources above it, are considered to undergo a starburst phase. We find that sources with \pam ~$>$ 43 tend to have higher specific star formation rates when compared to main sequence star forming galaxies at this redshift range, and populate the starbursts region. This result is coherent with the picture where these sources are compact starbursts with high sSFR and where (like local ULIRGs) a strong silicate absorption feature at $9.7\,\mu$m is present in their mid-IR spectra.

\section{Discussion}
Previous studies predicted that the number of such silicate break sources 
could reach $\sim$ 900 to 1500 sources per square degree, depending on the assumed model (e.g. Tagagi \& Pearson 2005). Furthermore, Kasliwal et al. (2005), based on $16\,\mu$m IRS data, reported that such sources account for more than half of all galaxies at $z$ $\sim$ 1$-$2 predicted by various models.  The samples of silicate-break galaxies and MIPS dropout sources that we have found here can place strong limits on the number of 1.0 $<$ $z$ $<$ 2.0 infrared luminous galaxies with similar properties.

In the 250 arcmin$^{2}$ covered by PACS in the two GOODS fields, we have detected $\sim$30 (7) silicate absorbed candidates at 1.0 $<$ $z$ $<$ 2.0 with \pam ~$>$ 43  and \lir ~$>$ ~10$^{11}$ \lsol ~(\lir ~$>$ ~10$^{12}$ \lsol). Assuming that all sources with \pam ~$>$ 43 at this redshift are silicate absorbed starbursts, and neglecting the effects of cosmic variance, this implies a surface density of  $\sim$ 540 (220) sources deg$^{-2}$, significantly lower than previous predictions. Using the co-moving volume within 1 $<$ $z$ $<$ 2, this implies a space density of Si absorbed ULIRGs in this redshift bin of $\sim$ 2.0 ($\pm$ 0.3) $\times$ 10$^{-5}$ Mpc$^{-3}$. Comparing the numbers with the rest high-$z$ sample of GOODS-H, we find that these 
sources account for the $\sim$ 8\% and 16\% of the \lir ~$>$ ~10$^{11}$ \lsol ~and \lir ~$>$ ~10$^{12}$ \lsol ~sources at 1.0 $<$ $z$ $<$ 2.0.  We note though, that our sample is not 
complete. As illustrated in Fig. 9, sources with silicate absorption feature but cold $T_{\rm dust}$ have \pam ~$<$ 43 and are missed from the selection. We therefore conclude that this estimate is a rather conservative lower limit. Moving to the dropout sample, we have identified 11 sources with \lir ~$>$ ~10$^{11}$ \lsol, accounting for approximately 1$-$2\% of the population of 1 $<$ $z$ $<$ 2 infrared luminous galaxies in GOODS-H as a whole. We note that by comparing to the whole GOODS-H sample in the same redshift range, all effects of incompleteness are taken into account. Finally, these number should be treated as upper limits, given that some of these could be spurious detections, as discussed in Section 2. 
\begin{figure*}
\centering
\includegraphics[scale=0.39]{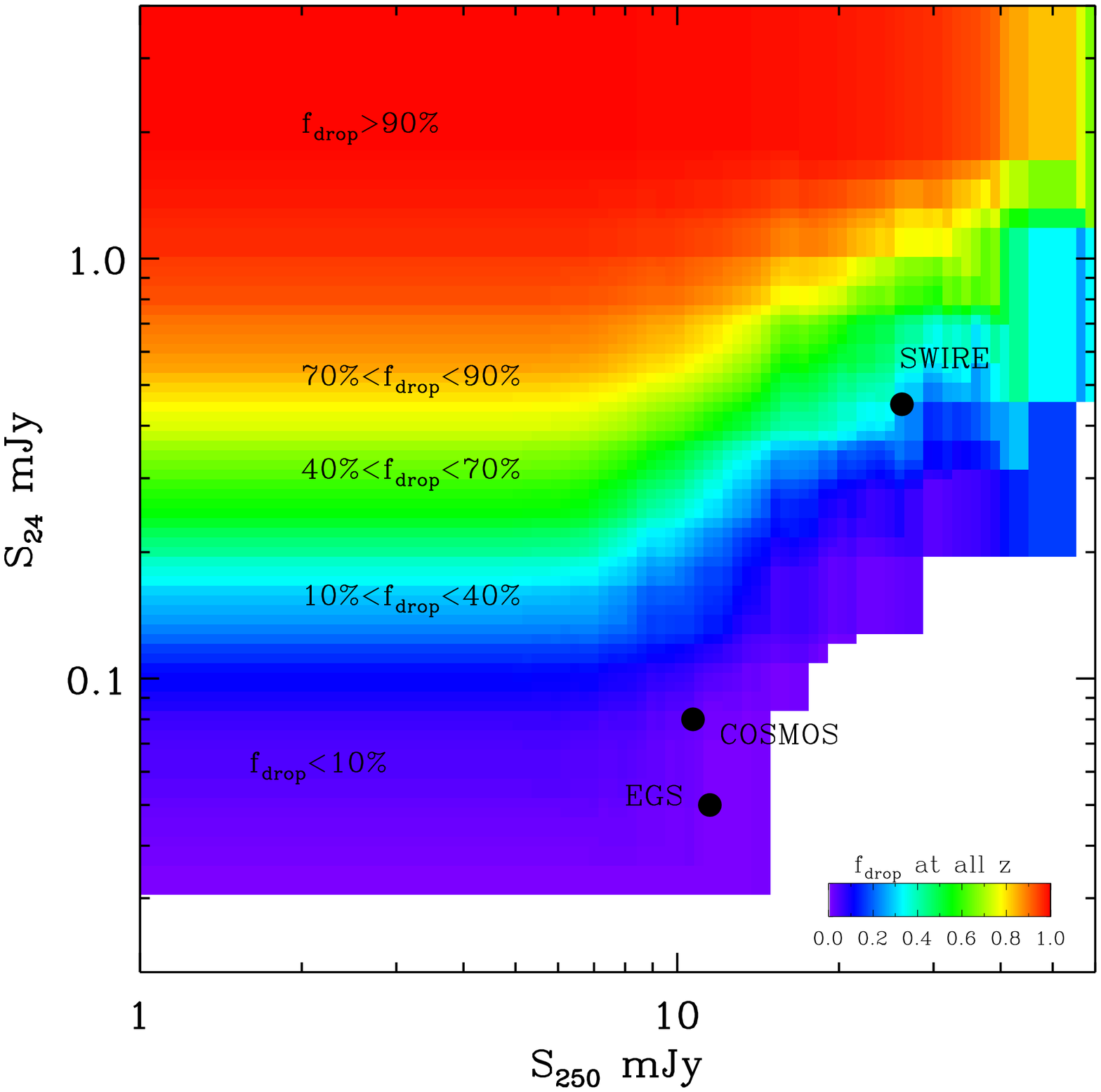}
\includegraphics[scale=0.39]{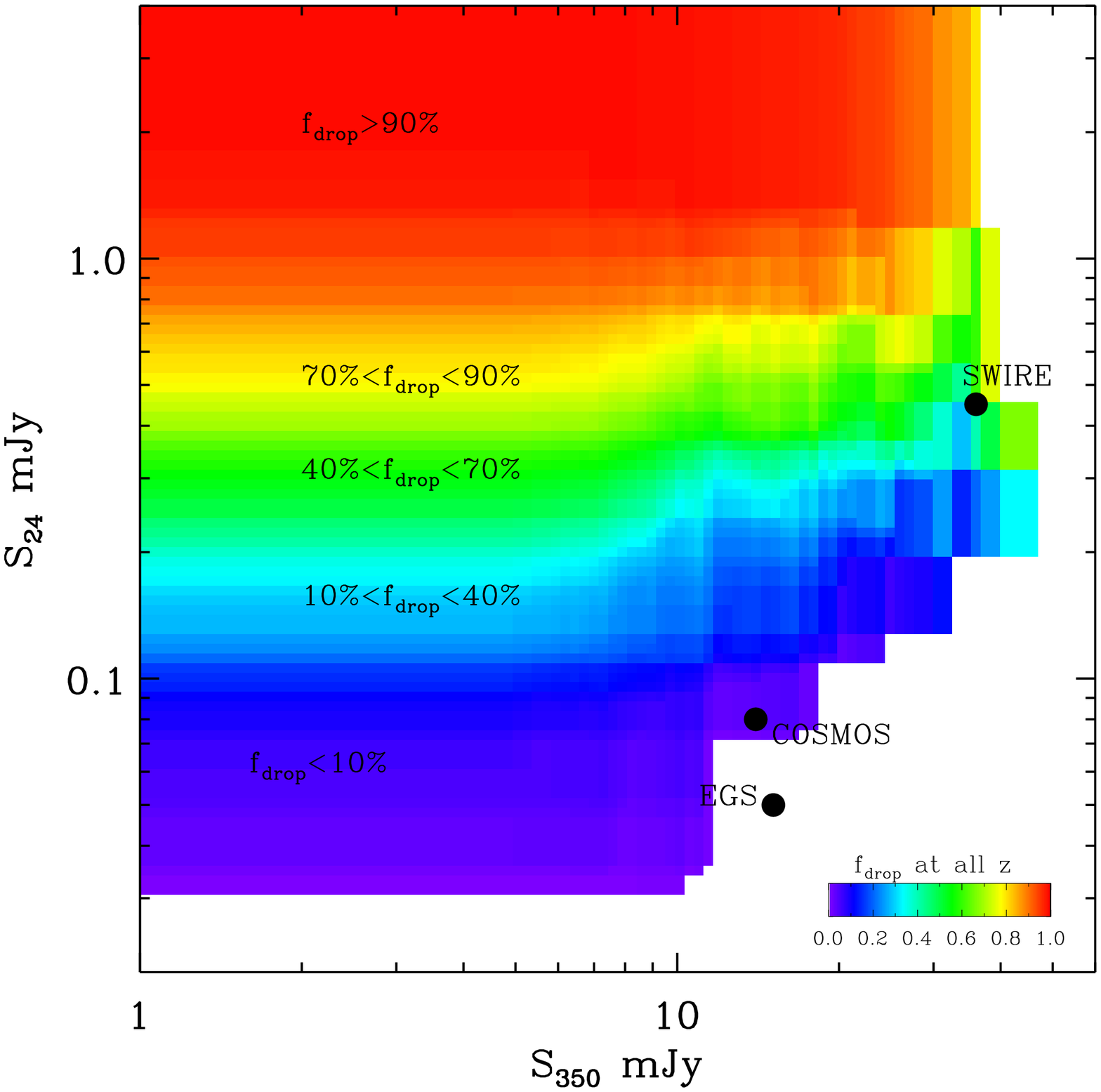}
\caption{The fraction of sources expected to be missed by the $24\,\mu$m band  at all redshifts as a function of the 24$\mu$m vs $250\,\mu$m and the $24\,\mu$m vs to $350\,\mu$m depth. Black circles indicate the depths of PACS and MIPS$24\,\mu$m surveys for some of the most important cosmological fields, i.e EGS, SWIRE and COSMOS.}
\label{FigGam}%
\end{figure*}

These estimates confirm that for the GOODS surveys, $24\,\mu$m observations recover the vast majority of $z$ $<$ 2 sources and do not introduce a strong selection bias.  Furthermore, we confirm that using the $24\,\mu$m catalogues to define priors for the extraction of PACS sources should only miss a small fraction of high-z sources. Of course the ratio between the number of sources missed by the $24\,\mu$m band and the total number of sources in the PACS bands strongly depends on the depth of the MIPS $24\,\mu$m and PACS observations. This is illustrated in Fig. 14, where we show the fraction of the expected dropouts at 0.2 $<$ $z$ $<$ 0.6 and 1.0 $<$ $z$ $<$ 1.7  as a function of the 24- to $100\,\mu$m and the 24- to $160\,\mu$m depth. These estimates are solely based on observations (i.e. on the GOODS-H sample), without any further assumptions. We simply calculate the ratio of $f_{\rm drop}$: \\

$\begin{array}{c}{\rm number ~of ~sources ~with ~S_{\rm 100} ~>~ \alpha ~\& ~S_{\rm 24} ~<~ \beta} \\ \overline{\rm number ~of ~sources ~with 
~S_{\rm 100} ~> ~\alpha ~\& ~S_{\rm 24} ~> ~\beta} 
\end{array} $
\\

\noindent where $\alpha$ and $\beta$ are free parameters. These diagrams can be used as diagnostic of the fraction of sources that 
will be missed by a source extraction method based on $24\,\mu$m priors for several extragalactic \h ~surveys, e.g. PEP (Lutz et al. 2011), HerMES (Oliver et al. in prep).  We conclude that for the major extragalactic surveys,  
there is not a large population of silicate break galaxies that would have been undetected in $24\,\mu$m \s ~data as some authors have previously suggested. This discrepancy could be indicative of an evolution of the strength of the  silicate absorption feature with time. If the silicate features were weaker in early galaxies due to more extended star forming regions, then theoretical models based on local templates would grossly overestimate the number density of such sources. This scenario is in line with recent findings, 
where the star formation activity of the majority of high$-z$ galaxies (including those with high \lir), 
is similar to that of local spiral galaxies (e.g. Daddi et al. 2010, Genzel et al. 2010).   

Finally, although in this work we have mainly focused on PACS data, it is worth attempting  to quantify the fraction of sources that would 
be missed by a $24\,\mu$m prior based  source extraction in the SPIRE bands. The philosophy behind Fig. 15 is identical to that of Fig. 14, but this time considering the 
whole redshift range of the sources and the $S_{\rm 250}$ (left) and $S_{\rm 350}$ (right) flux density limits. 
We note that these detection limits, indicate the expected instrumental noise and do not take into account the confusion noise, which is dominant in the SPIRE bands.   
 
\section{Conclusions}
We have presented a study dedicated to sources that exhibit atypical \pam ~colours using the deepest PACS data to date, obtained as part of the GOODS-H program. By performing blind source extraction we searched for sources that are bright in the far-IR but undetected at $24\,\mu$m, i.e. for $24\,\mu$m dropout galaxies. Then we investigated the properties of sources in the GOODS-H sample with red \pam ~colours searching for the population of silicate-break galaxies. The main results are summarized below:
 
\begin{itemize}
\item We have identified 21 PACS sources that are undetected at $24\,\mu$m (down to a 3$\sigma$ detection limit of $\sim$ 20 $\mu$Jy). These $24\,\mu$m dropout sources are found to have a bimodal redshift distribution, with peaks centred at $z$ $\sim$ 0.4 and $\sim$ 1.3, and are expected to exhibit strong silicate absorption features, responsible for their depressed $24\,\mu$m emission. Among the sources in higher redshift we identify 10 LIRGs and one ULIRG at $z$ $=$1.68. This enables us to place upper limits in the fraction of LIRGs/ULIRGs that are missed by $24\,\mu$m surveys. 
 \item The vast majority of \h ~PACS sources are detected at $24\,\mu$m, indicating that a prior$-$based source extraction based on the $24\,\mu$m emission of the galaxies suffers only very modest incompleteness, with MIPS dropout sources accounting only for $\sim$ 2\% of the infrared luminous population in the GOODS fields. Although this fraction is negligible for the GOODS surveys, $24\,\mu$m dropouts may be a concern for other \h ~extragalactic surveys with shallower $24\,\mu$m data.
\item Based on the mid- and far-IR colours of sources in the GOODS-H sample, we demonstrated that sources with \pam ~$>$ 43 and \iri ~$>$ 4 are located in a narrow redshift bin, $1.0$ $<$ $z$ $<$ $1.7$. Furthermore, we provided evidence that sources selected in this manner are starburst$-$dominated and with compact geometries. Similarly to the dropouts, the red \pam ~colours of these sources are attributed to the  $9.7\,\mu$m silicate absorption feature in their mid-IR spectra that enters into the $24\,\mu$m band.  We  characterize them  as silicate-absorbed galaxies.
\item The infrared luminosity of these silicate-absorbed galaxies, when derived based on their monochromatic $24\,\mu$m flux density, is on average underestimated by a factor $\sim$ 3. They account for about 16\% of the ULIRGs in the GOODS fields with a space density of 2.0 $\times$ 10$^{-5}$ Mpc$^{-3}$.
\item We provide diagnostic diagrams to estimate the fraction of sources expected to be missed in the $24\,\mu$m band for several \h ~extragalactic surveys, and predict that for most of them that fraction is less that 10\%. 
\end{itemize}

\begin{acknowledgements}
G.E.M. acknowledge the support of the Centre National de la Research Scientifique (CNRS) and the University of Oxford.
H.S.H and D.E. acknowledge the support of the Centre National d'Etudes Spatiales (CNES)
PACS has been developed by a consortium of institutes led by MPE (Germany) and including 
UVIE (Austria); KU Leuven, CSL, IMEC (Belgium); CEA, LAM (France); MPIA (Germany);
INAFIFSI/OAA/OAP/OAT, LENS, SISSA (Italy) and IAC (Spain). This development has been supported by 
the funding agencies BMVIT (Austria), ESA-PRODEX (Belgium), CEA/CNES (France), DLR (Germany),
ASI/INAF (Italy), and CICYT/MCYT (Spain). 
SPIRE has been developed by a consortium of institutes led by Cardiff University (UK) and 
including Univ. Lethbridge (Canada); NAOC (China); CEA, LAM (France); IFSI, Univ. Padova (Italy); 
IAC (Spain); SNSB (Sweden); Imperial College London, RAL, UCL-MSSL, UKATC, 
Univ. Sussex (UK); and Caltech, JPL, NHSC, Univ. Colorado (USA). This development has been 
supported by national funding agencies: CSA (Canada); NAOC (China); CEA, CNES, CNRS (France); 
ASI (Italy); MCINN (Spain); Stockholm Observatory (Sweden); STFC (UK); and NASA (USA). This work is based [in part] on observations made with Herschel, a European Space Agency Cornerstone Mission with significant participation by NASA. Support for this work was provided by NASA through an award issued by JPL/Caltech.
\end{acknowledgements}

\end{document}